\documentclass[preprintnumbers,aps,prd,floatfix,nofootinbib,onecolumn]{revtex4}
\usepackage{qcd}
\usepackage{epsfig}
\usepackage{graphicx}
\usepackage[normalem]{ulem}
\usepackage[pagebackref]{hyperref}
\usepackage{bm}
\usepackage{float}
\usepackage{orcidlink}
\usepackage{etoolbox}

\definecolor{nicegreen}{rgb}{0.,0.5,0.}

\begin{document}

\title{QCD factorization with multihadron fragmentation functions}

\preprint{JLAB-THY-24-4246}

\author{T.~C.~Rogers \orcidlink{0000-0002-0762-0275}\,}
\email{trogers@odu.edu}
\affiliation{Department of Physics, Old Dominion University, Norfolk, VA 23529, USA}
\affiliation{Jefferson Lab, 12000 Jefferson Avenue, Newport News, VA 23606, USA}
\author{M.~Radici \orcidlink{0000-0002-4542-9797}\,}
\email{marco.radici@pv.infn.it}
\affiliation{INFN - Sezione di Pavia, via Bassi 6, I-27100 Pavia, Italy}

\author{A.~Courtoy\orcidlink{0000-0001-8906-2440}\,}
\email{aurore@fisica.unam.mx}
\affiliation{Instituto de F\'isica,
  Universidad Nacional Aut\'onoma de M\'exico, Apartado Postal 20-364,
  01000 Ciudad de M\'exico, Mexico}

\author{T.~Rainaldi \orcidlink{0000-0002-8342-6765}\,}
\email{train005@odu.edu}
\affiliation{Department of Physics, Old Dominion University, Norfolk, VA 23529, USA}

\begin{abstract}
Important aspects of QCD factorization theorems are the properties of the objects involved that can be identified as universal. 
One example is that the definitions of parton densities and fragmentation functions for different types of hadrons differ only in the identity of the nonperturbative states that form the matrix elements, but are otherwise the same.  This leads to independence of perturbative calculations on nonperturbative details of external states. It also lends support to interpretations of correlation functions as encapsulations of intrinsic nonperturbative properties. These characteristics have usually been presumed to still hold true in fragmentation functions even when the observed nonperturbative state is a small-mass cluster of $n$ hadrons rather than simply a single isolated hadron.  
However, the multidifferential aspect of cross sections that rely on these latter types of fragmentation functions complicates the treatment of kinematical approximations in factorization derivations. That has led to recent claims that the operator definitions for fragmentation functions need to be modified from the single hadron case with nonuniversal prefactors.  
With such concerns as our motivation, we retrace the steps for factorizing the unpolarized semi-inclusive $e^+e^-$ annihilation cross section and confirm that they do apply without modification to the case of a small-mass multihadron observed in the final state. In particular, we verify that the standard operator definition from single hadron fragmentation, with its usual prefactor, remains equally valid for the small-mass $n$-hadron case with the same hard parts and evolution kernels, whereas the more recently proposed definitions with nonuniversal prefactors do not. 
Our results reaffirm the reliability of most past phenomenological applications of dihadron fragmentation functions. 
\end{abstract}

\date{December 16, 2024}

\maketitle

\section{Introduction}
\label{s.intro}

Factorizable high-energy collisions 
in QCD are describable in terms of convolution products of a perturbatively calculable factor
with
one or more factors that encapsulate the description of nonperturbative hadronic properties. 
The nonperturbative factors can take the form of parton distribution functions, which describe the distribution of partons in a target, or fragmentation functions, which characterize the hadronization of a parton into particular types of final states.
In the simplest cases where processes are very inclusive, the parton distribution and fragmentation functions depend on only one variable, the momentum fraction of the parent particle involved, in addition to an auxiliary scale associated with the renormalization group. For such cases, collinear factorization is the relevant theoretical framework, and the derivations lead to definitions for universal nonperturbative correlation functions like parton distribution or fragmentation functions. 
A notable advantage of collinear factorization is that it allows one to exploit the simplifications arising from inclusive sums over final states.

The fragmentation of a parton into a single hadron can be extended to cases for which a parton fragments into a small-mass cluster of $n$ hadrons (which we will call an ``$n$-hadron'').
By incorporating $n$-hadron fragmentation functions within collinear phenomenology,  one is able to exploit the advantages of collinear factorization to study quantities like chiral-odd distributions (transversity, twist-3 $e(x)$) and related moments (tensor and scalar charge) without needing to resort to more complicated QCD formalisms like transverse momentum dependent (TMD) factorization, {\it e.g.}~\cite{Gamberg:2022kdb}, or generalized parton distribution frameworks, {\it e.g.}~\cite{Goldstein:2013gra}. 
As such, there is an long history of projects dedicated to 
extracting dihadron fragmentation functions phenomenologically~\cite{Boer:2003ya,Courtoy:2012ry,Radici:2015mwa, Matevosyan:2018icf} to access objects like the transversity distribution~\cite{Collins:1993kq,Collins:1994ax,Jaffe:1998hf,Bianconi:1999cd,Radici:2001na,Bacchetta:2002ux,Bacchetta:2003vn,Bacchetta:2004it,Bacchetta:2011ip,Bacchetta:2012ty,Radici:2015mwa,Bacchetta:2008wb, Courtoy:2015haa, Radici:2016lam,Radici:2018iag,Benel:2019mcq} and interpreting them nonperturbatively. 

Ultimately, connecting phenomenological extractions to insights about nonperturbative structures requires precisely defined QCD operator matrix elements for the $n$-hadron fragmentation functions. Given that the steps of factorization derivations with fragmentation functions appear to be independent of the identity of the observed hadronic state, it has usually been presupposed that the standard fragmentation function definition that has long been used for single hadron fragmentation applies equally well to the $n$-hadron case, with the only change necessary being that the single hadron state is replaced by an $n$-hadron state in the matrix element (see, for example, Sec.~12.13.3 of Ref.~\cite{Collins:2011qcdbook}).
Early work on dihadron fragmentation functions like Ref.~\cite{Collins:1993kq,Collins:1994ax} also used a dihadron fragmentation function definition that exactly matched the single hadron case (compare Eq.~(10) of~\cite{Collins:1993kq} with Eq.~(2.14) of \cite{Collins:1994ax}\footnote{The fragmentation functions in Refs.~\cite{Collins:1993kq,Collins:1994ax} are actually a factor of $z$ times what is normally called a fragmentation function, see Ref.~\cite{Collins:1981uw}.}). The standard fragmentation function definition has also motivated most of the phenomenological work performed over the past few decades like, for example, Eqs.~(9) and~(11) of Ref.~\cite{Bianconi:1999cd} (see also Ref.~\cite{Matevosyan:2018icf}). Finally, it is the definition used for nonperturbative modeling of dihadron fragmentation functions in works such as Refs.~\cite{Bianconi:1999uc,Bacchetta:2006un,Matevosyan:2013aka,Matevosyan:2017alv,Matevosyan:2017uls}. In the future, having a clear and unambiguous definition for the fragmentation functions will be important for the expanding efforts being put into understanding nonperturbative physics through the extraction and theoretical analysis of partonic correlation functions. 

However, a recent study in Ref.~\cite{Pitonyak:2023gjx} argues that the definition of a dihadron (or $n$-hadron) fragmentation function needs to be modified with momentum-fraction-dependent overall prefactors, thus calling into question the validity of earlier work with dihadron fragmentation functions.
The authors of Ref.~\cite{Pitonyak:2023gjx} further argue that the standard definitions of dihadron fragmentation functions used in most of the past phenomenological efforts, like those based on Ref.~\cite{Bianconi:1999cd,Bacchetta:2002ux,Boer:2003ya} and models like Refs.~\cite{Bianconi:1999uc,Bacchetta:2006un,Matevosyan:2013aka,Matevosyan:2017alv,Matevosyan:2017uls}, lack a robust interpretation within a parton model framework. They propose a new definition, which has since been adopted as the theoretical basis for phenomenological applications by the JAM collaboration~\cite{Cocuzza:2023vqs,Cocuzza:2023oam}.

In this paper, we show that the standard definition, with its usual prefactor, is actually the correct one in that it is the universal operator definition that is consistent with ordinary collinear factorization with the usual hard parts and evolution kernels. We also discuss how the parton model interpretation of the standard dihadron fragmentation function definition generalizes directly to the multihadron case. 
Fortunately, therefore, earlier work with dihadron fragmentation functions based on the standard definition and its interpretation is not compromised.

As an example process, we will consider semi-inclusive $e^+ e^-$ annihilation (SIA), as this is a simple and typical application of fragmentation functions that exposes the advantages of the collinear factorization framework as compared with the TMD one~\cite{Bacchetta:2015ora}. 
Experimental data for dihadron production in this process have been produced at Belle, both for the Artru-Collins asymmetry~\cite{Vossen:2011fk} and, more recently, dihadron multiplicities \cite{Belle:2017rwm}.

The paper is organized as follows: In Sec.~\ref{s.setup}, we introduce the notation and the conventions/approximations used to describe the process of lepton-antilepton annihilation into a final hadronic state. In Sec.~\ref{s.kin}, we describe the kinematics and the observables (cross section, structure functions) for this process. In Sec.~\ref{s.fact}, we highlight the main steps to achieve a factorized form of the hadronic tensor. In Sec.~\ref{s.ffs}, we scrutinize the operator definition of fragmentation function that is consistent with factorization. In Sec.~\ref{s.singlehad}, we recover the familiar expression for the single-hadron $e^+ e^-$ cross section consistently with the factorized expression of the hadronic tensor. Similarly, in Sec.~\ref{s.dihadronexample} we recover the standard expression of the $e^+ e^-$ cross section for inclusive dihadron production adopted in many phenomenological works, using two different sets of dihadron internal variables as illustrative examples. In Sec.~\ref{s.alternative}, we comment on the alternative operator definition of multi-hadron fragmentation function presented in Ref.~\cite{Pitonyak:2023gjx}. Finally, in Sec.~\ref{s.end} we summarize our main observations and make concluding remarks.

\section{Setup}
\label{s.setup}

We focus on the treatment of factorization for the semi-inclusive annihilation (SIA) process,
\begin{equation}
\label{e.process}
l^+ l^- \to \gamma^* \to \left\{ h_1, h_2,\dots,h_n \right\} + X  \, .
\end{equation}
The virtual photon has four-momentum $q$, and $q^2 = Q^2$. The ``$\left\{ h_1, h_2,\dots,h_n \right\}$'' denotes a high energy, small-mass cluster of $n$ hadrons whose momenta we label $\hv{1}, \hv{2},\dots, \hv{n}$. We will call such a cluster of hadrons an ``$n$-hadron,'' or just a dihadron in the special case of $n = 2$. The total momentum of the $n$-hadron is
\begin{equation}
p_h = \sum_{i=1}^n \hv{i} \, .
\end{equation}
We define the two invariants $p_h^2 = M_h^2$ and $z = 2 p_h \cdot q/q^2$. 
In this paper, we will restrict to the region $M_h^2 \ll Q^2$ and we will only consider
channels where one parton emitted from the hard scattering fragments into a small $M_h^2$ cluster of $n$ hadrons at fixed $z$. We also require that $|z_i - z_j|$ is limited from above for any pair $i,j = 1,..,n$, with $z_i = 2 p_{h_i} \cdot q/q^2$, in order to avoid power-suppressed (p.s.) contributions from soft partons with large relative momentum which could become important and break factorization in practical situations, particularly when gluon densities are large (see Refs.~\cite{Qiu:1990xxa,Qiu:2001hj,Qiu:2003cg} for examples in the Drell-Yan process).

We will retrace the steps for arriving at the collinear factorization formula,
\begin{equation}
\label{e.fact_formula}
\frac{\diff{\sigma}{}}{\diff{Y}{}} = \int_z^1 \frac{\diff{\zetasym}{}}{\zetasym^2} \parz{2 E_{\hat{k}} (2 \pi)^3 \frac{\diff{\hat{\sigma}}{}}{\diff[3]{\vect{\hat{k}}}{}} } d(\zetasym,\left\{p_h\right\}) + \text{p.s.} \, ,
\end{equation}
starting from the unfactorized cross section, with the usual Lorentz invariant phase space 
\begin{equation}
\label{e.phase_space}
 \diff{Y}{} \equiv \frac{\diff[3]{\vect{p}_{h_1}}{}}{2 E_{\hv{1}} (2 \pi)^3} \frac{\diff[3]{\vect{p}_{h_2}}{}}{2 E_{\hv{2}} (2 \pi)^3} \times \cdots \times \frac{\diff[3]{\vect{p}_{h_n}}{}}{2 E_{\hv{n}} (2 \pi)^3} \, .
\end{equation}
Equation~\eqref{e.fact_formula} is the basic semi-inclusive annhilation factorization formula in, for example, Eq.(12.13) of Ref.~\cite{Collins:2011qcdbook} but now generalized to the case of an $n$-hadron. 
Here, $\vect{\hat{k}}$ is the 3-momentum of the fragmenting parton and $\zetasym$ is the light-cone momentum fraction $$\zetasym = p_h^+/k^+$$ in a hadron frame (see below for a review of frames).  
The $2 E_{\hat{k}} (2 \pi)^3 \diff{\hat{\sigma}}{}/\diff[3]{\vect{\hat{k}}}{}$ in parentheses is the semi-inclusive cross section at the partonic level for producing a parton with momentum $\vect{\hat{k}}$ and energy $E_{\hat{k}}$, and $d(\zetasym,\left\{p_h\right\})$ is a fragmentation function for a system of $n$ hadrons (an ``$n$-hadron'') carrying a total fraction $\zetasym$ of the initial parton's momentum. The $\left\{ p_h \right\}$ symbolizes dependence on all of the observed $n$ hadron momenta, 
\begin{equation}
\left\{p_h\right\} \equiv \left\{\hv{1},\hv{2},\dots,\hv{n} \right\} \, .
\label{e.multihad_p}
\end{equation}
The ``$\text{p.s.}$'' means ``power suppressed'' and refers to terms in \eref{fact_formula} that vanish like powers of $\Lambda^2/Q^2$ as $Q \to \infty$, where $\Lambda$ is a generic small-mass scale like $\Lambda_{\mathrm{QCD}}$ or a small hadron mass. 

In the special case of a single observed hadron ($n=1$), 
\begin{equation}
\diff{Y}{} \to \frac{\diff[3]{\vect{p}_h}{}}{2 E_p (2 \pi)^3} \,  \qquad \text{and} \qquad  d(\zetasym,\left\{p_h\right\}) \to d(\zetasym,p_h) \, ,
\end{equation}
with $d(\zetasym,p_h)$ matching the standard definition of a single-hadron fragmentation function for an observed hadron with total momentum $p_h$. 

General derivations of \eref{fact_formula} in QCD are well established for the single hadron case (see chapter 12 of Ref.~\cite{Collins:2011qcdbook} and references therein), and the steps carry over without modification to the case of an observed $n$-hadron. The task of this paper, therefore, only amounts to examining whether any changes might be necessary to account for the kinematics of an $n$-particle final state phase space on the left-hand side.
Therefore, we will work with the following simplifications to streamline the discussion:
\begin{enumerate}
\item For our purposes, it will be enough to work with non-gauge theories and to consider a theory of quarks with no color and only one flavor. We will take all quarks to have unit fractional charge, $e_q = 1$. A sum over quarks and antiquarks will be implied and not written out explicitly.
\item We will consider only the leading regions that contribute to the semi-inclusive cross section at zeroth order in the hard part. That is, a massless, on-shell quark-antiquark pair is produced with a single parton fragmenting into an $n$-hadron. (See \fref{LeadingRegion} below.)
\item We will assume that there is only one flavor, $h$, of hadron in the final state, whose mass is $m$.
\item All expressions will be given in exactly four dimensions rather than with dimensional regularization made explicit.  
\item When we encounter ultraviolet divergent transverse momentum integrals, we will assume that regulators have been applied, but we will not indicate it explicitly in the notation. We will suppress explicit dependence on auxiliary variables like the renormalization group scale $\mu$ except in places where it is necessary for the discussion. 
\end{enumerate}
These simplifications are already reflected in our expression of \eref{fact_formula} above, e.g. in the lack of sums over parton flavor. Once the main issues related to kinematical approximations have been clarified, the extension to full QCD, multiple flavors, etc, will be straightforward, based on existing derivations.  

Also, we stress that for this paper we focus only on the contribution from the single parton fragmentation channel. In applications, contributions where separate partons fragment independently can be important, even at relatively low scales. However, these involve factorization with single hadron fragmentation functions. Such treatments can be combined, as is typically done, with a multihadron fragmentation function formalism in additional steps.  

\section{Kinematics and observables}
\label{s.kin}

\subsection{Cross sections} 
\label{s.crosssection}

A differential element of the cross section (including any spin dependence) for $N_\text{tot}$ final state particles, neglecting the lepton masses, is
\begin{equation}
\diff{\sigma} = \frac{|M|^2}{2 Q^2} \frac{\diff[3]{\vect{p}_{1}}}{2 E_1 (2 \pi)^3} \frac{\diff[3]{\vect{p}_2}}{2 E_2 (2 \pi)^3} \times \cdots \times (2 \pi)^4 \delta^{(4)}\parz{q - p_X} \, ,
\end{equation}
where $p_X = \sum_i^{N_{\mathrm{tot}}} p_i$, and $M$ is the scattering amplitude. The total cross section averaged over lepton spins and separated into leptonic and hadronic tensors is 
\begin{equation}
\label{e.cross}
\sigma = \frac{4 \pi e^4}{2 Q^6} L^{\mu \nu} W_{\mu \nu} \, \qquad \text{with} \qquad L^{\mu \nu} = l^\mu l'^\nu + l'^\mu l^\nu - l \cdot l' g^{\mu \nu}
\end{equation}
and 
\begin{equation}
\label{e.hadtens}
W^{\mu \nu} = \frac{1}{4 \pi} \SumInt_X \langle 0 | j^\mu(0) |X\rangle \langle X| j^\nu(0) | 0 \rangle (2 \pi)^4 \delta^{(4)} \parz{q - p_X} = \frac{1}{4 \pi} \int \diff[4]{x} e^{i q \cdot x} \langle 0 | j^\mu(x) j^\nu(0) | 0 \rangle\, ,
\end{equation}
with
\begin{equation}
\label{e.finalstates}
\SumInt_X |X\rangle \langle X| \equiv 
\sum_{\overset{\text{All \#'s}}{\text{of particles}}} \int \frac{\diff[3]{\vect{p}_1}}{2 E_1 (2 \pi)^3} \frac{\diff[3]{\vect{p}_2}}{2 E_2 (2 \pi)^3} \times \cdots | p_1, p_2, \dots \rangle \langle p_1, p_2, \dots |\, ,
\end{equation}
using the standard normalization for Heisenberg ``out'' eigenstates. The $L^{\mu \nu}$ is the spin averaged leptonic tensor for leptons with momentum $l$ and $l'$.

For the $n$-particle semi-inclusive case in \eref{process}, one makes the replacement
\begin{align}
\label{e.hadtensSI}
W^{\mu \nu} \to 
&{} \frac{1}{4 \pi} \SumInt_X \int_{V} \frac{\diff[3]{\vect{p}_{h_1}}}{2 E_{h_1} (2 \pi)^3} \int_{V} \frac{\diff[3]{\vect{p}_{h_2}}}{2 E_{h_2} (2 \pi)^3} \times \cdots \times \int_{V} \frac{\diff[3]{\vect{p}_{h_n}}}{2 E_{h_n} (2 \pi)^3} \no
&{} \qquad \times \langle 0 | j^\mu(0) a^\dagger_{\vect{p}_{h_n}} \cdots a^\dagger_{\vect{p}_{h_2}} a^\dagger_{\vect{p}_{h_1}} |X\rangle \langle X| a_{\vect{p}_{h_1}} a_{\vect{p}_{h_2}} \cdots a_{\vect{p}_{h_n}} j^\nu(0) | 0 \rangle (2 \pi)^4 \delta^{(4)} \parz{q - p_{X} - p_h} \no
= &{}  \frac{1}{4 \pi} \SumInt_{X} \int_{V} \diff{Y} \; \langle 0 | j^\mu(0) |\hv{1},\hv{2},\dots,\hv{n},X\rangle \langle \hv{1},\hv{2},\dots,\hv{n}, X| j^\nu(0) | 0 \rangle (2 \pi)^4 \delta^{(4)} \parz{q - p_{X} - p_h} \, .
\end{align}
Now the momenta $p_1, p_2, \dots$ without an ``$h$'' subscript in \eref{finalstates} label the 3-momenta of hadrons that are not counted as part of the $n$-hadron.
For factorization to hold,
the range of integration for each final state momentum will need to be restricted to some appropriate momentum space volume, as indicated by the $V$ subscript on the integration symbols.  
On the last line, we have used the abbreviation
\begin{equation}
\label{e.basicps}
\int_{V} \diff{Y} (\cdots) = \int_{V} \frac{\diff[3]{\vect{p}_{h_1}}}{2 E_{h_1} (2 \pi)^3} \int_{V} \frac{\diff[3]{\vect{p}_{h_2}}}{2 E_{h_2} (2 \pi)^3} \times \cdots \times \int_{V} \frac{\diff[3]{\vect{p}_{h_n}}}{2 E_{h_n} (2 \pi)^3} (\cdots) \; .
\end{equation}
For the semi-inclusive differential cross section, the $\int_V$ integrals are dropped and $\diff{Y}$ is moved to the left-hand side. We define
\begin{equation}
W^{\mu \nu}_{\text{SI}} \equiv \langle N_n \rangle \frac{\diff{W^{\mu \nu}}}{\diff{Y}{}} \equiv
\frac{1}{4 \pi} \SumInt_{X}\; \langle 0 | j^\mu(0) |\hv{1},\hv{2},\dots,\hv{n},X\rangle \langle \hv{1},\hv{2},\dots,\hv{n}, X| j^\nu(0) | 0 \rangle (2 \pi)^4 \delta^{(4)} \parz{q - p_{X} - p_h} \, \label{e.WSI}
\end{equation}
where
\begin{equation}
\label{e.inclusive}
\int \diff{Y}{} W^{\mu \nu}_{\text{SI}}= \langle N_n \rangle W^{\mu \nu} \, 
\end{equation}
follows from the definition of the inclusive cross section, 
and $\langle N_n \rangle $ is the total multiplicity of the $n$-hadrons (for example, for $n=2$, $\langle N_2 \rangle$ is the number of dihadron pairs).  For the semi-inclusive case,~\eref{cross} then becomes 
\begin{equation}
\frac{\diff{\sigma}}{\diff{Y}}  = \frac{4 \pi e^4}{2 Q^6} L_{\mu \nu} W^{\mu \nu}_{\text{SI}} \, . \label{e.strucexp}
\end{equation}
For $n=1$, 
\begin{equation}
E_{p_h} \frac{\diff{\sigma}}{\diff[3]{\vect{p}_h}{}} = \frac{2 \alpha^2}{Q^6} L_{\mu \nu} W^{\mu \nu}_{\text{SI},n=1} \, . \label{e.tensordecomp}
\end{equation}
In the special case that the produced particle is a single massless and on-shell parton with momentum $\vect{\hat{k}}$, \eref{tensordecomp} is 
\begin{equation}
E_{\vect{\hat{k}}} \frac{\diff{\hat{\sigma}}}{\diff[3]{\vect{\hat{k}}}{}} = \frac{2 \alpha^2}{Q^6} L_{\mu \nu} \widehat{W}^{\mu \nu}_{\text{SI}} \, , \label{e.tensordecomppartonic}
\end{equation}
where $\widehat{W}^{\mu \nu}_{\text{SI}}$ is exactly \eref{WSI} but with a single partonic final state rather than the $n$-hadron $\left\{ p_h \right\}$. That is, 
\begin{equation}
\widehat{W}^{\mu \nu}_{\text{SI}}(\hat{z}) \equiv
\frac{1}{4 \pi} \SumInt_{X} \; \langle 0 | j^\mu(0) |\hat{k},X\rangle \langle \hat{k} , X| j^\nu(0) | 0 \rangle (2 \pi)^4 \delta^{(4)} (q - p_{X} - \hat{k}) \, , \label{e.WSIhat}
\end{equation}
where $\hat{z} \equiv 2 \hat{k} \cdot q / Q^2$. 
Rather than working directly with 
\eref{fact_formula}, it is convenient to derive the equivalent formula for hadronic tensors, 
\begin{equation}
\label{e.fact_formula2}
W^{\mu \nu}_{\text{SI}}  = \int \frac{\diff{\zetasym}{}}{\zetasym^2} \widehat{W}^{\mu \nu}_{\text{SI}}(\hat{z}) \, d(\zetasym,\left\{p_h\right\}) + \text{p.s.} \, .
\end{equation}

The total number of final state $n$-hadrons in an event with an overall total of $N_{\text{tot}}$ final state hadrons equals the number of ways of choosing $n$ hadrons multiplied by the $n!$ ways of relabeling them,
\begin{equation}
n! {N_{\text{tot}}  \choose n}  = N_{\text{tot}} (N_{\text{tot}}-1) (N_{\text{tot}}-2) \times \cdots \times (N_{\text{tot}}-n+1) \, .
\end{equation}
The combinatorial factor follows automatically from the $a^\dagger_{\vect{p}_{h}}$'s and $a_{\vect{p}_{h}}$'s in \eref{hadtensSI}. In this way, one may express the number of $n$-hadrons in terms of the total number of hadrons on the right-hand side of \eref{inclusive} by writing, for example, $\langle N_2 \rangle = \langle N_{\text{tot}} (N_{\text{tot}}-1) \rangle$, $\langle N_3 \rangle = \langle N_{\text{tot}} (N_{\text{tot}}-1) (N_{\text{tot}}-2) \rangle$, etc. Similar combinatorial factors were found in the context of discussions of fragmentation functions in Refs.~\cite{Majumder:2004br,Pitonyak:2023gjx}.

\subsection{Reference frames and leading regions}

\begin{figure}[h!]
\centering
\includegraphics[width=8cm]
{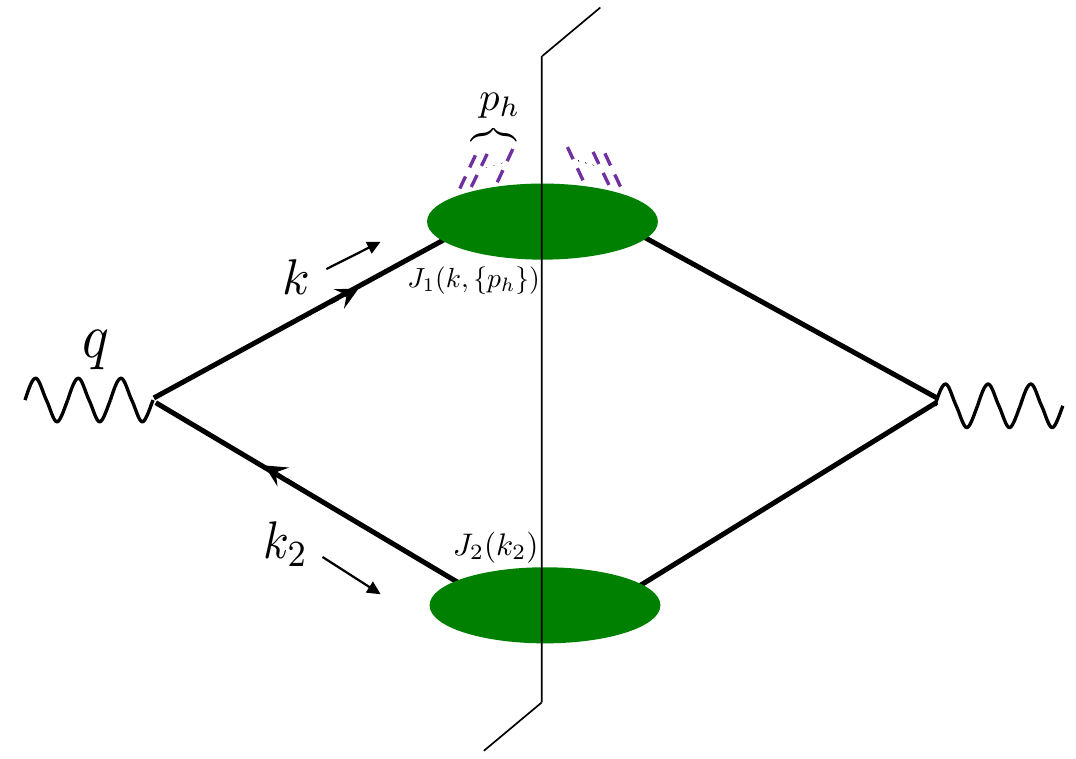}
\caption{The leading region contributing to $W^{\mu \nu}_{\text{SI}}$ at zeroth order in the hard part. The observed $n$-hadron is shown as the cluster of dashed lines emerging from the upper green subgraph. The momenta in the green bubbles are integrated over.}
\label{f.LeadingRegion}
\end{figure}

The relevant asymptotic limit for applying factorization is small $\Lambda_{\rm QCD}/Q$ with fixed $p_h^0/Q$ in the virtual photon center of mass system. Standard region analysis then gives as the dominant region (with zeroth order in the hard part) the structure shown in \fref{LeadingRegion}. Here, there are two jet regions represented by the green final state subgraphs labeled $J_1$ and $J_2$.  Lines inside the subgraphs are, respectively, collinear and anticollinear to $p_h$. The hard vertex has one outgoing parton with momentum $k$ that fragments into the observed $n$-hadron system with momentum $p_h$ and an unobserved subgraph $J_1$. The single other unobserved parton (momentum $k_2 = q - k$) hadronizes into an unobserved subgraph of hadrons\footnote{From Lorentz invariance, the components of parton momentum $k$ and the momenta in $\{p_h\}$ are not totally independent.} labeled $J(k_2)$. 

We recall the definition of kinematical variables (see Sec.~\ref{s.setup}), which will be useful later on,  
\begin{equation}
\label{e.zvars}
z \equiv \frac{2 p_h \cdot q}{Q^2} \, , \qquad \zn \equiv \frac{z}{2} + \sqrt{\frac{z^2}{4} - \frac{M_h^2}{Q^2}} = z + \order{\frac{M_h^2}{z Q^2}} \, .
\end{equation}
In the virtual photon center-of-mass, $z = 2 p_h^0/Q$.
So, the collinear factorization limit is $\Lambda_{\text{QCD}}/Q \to 0$ with fixed $z \approx \zn$. To obtain factorization, we will need fixed $M_h = \order{\Lambda_{\text{QCD}}}$.

In later sections, we will mostly work with three basic reference frames: a ``hadron frame,'' a ``lab frame,'' and a ``photon frame.'' The hadron frame is a center-of-mass frame where the observed hadron (or $n$-hadron) has zero transverse momentum. The momenta in this frame are labelled with an $H$ subscript,
\begin{align}
p_h {}&= \parz{p_{h,H}^+,\frac{M_h^2}{2 p_{h,H}^+},\T{0}{}}_H = \parz{\frac{\zn Q}{\sqrt{2}},\frac{M_h^2}{\zn Q \sqrt{2}}, \T{0}{}}_H \, , \qquad
q {}= \parz{q_H^+,\frac{Q^2}{2 q_H^+},\T{0}{}}_H \, = \parz{\frac{Q}{\sqrt{2}}, \frac{Q}{\sqrt{2}}, \T{0}{}}_H , \label{e.phlabel} \\
k &{}= \parz{k_H^+,k_H^-,\T{k}{H}}_H = \parz{p_{h,H}^+/\zetasym,k_H^-,\T{k}{H}}_H \, . \label{e.khlabel}
\end{align}
We also orient the positive $z$-axis along the direction of motion of $p_h$ so that $p_h^+ \geq p_h^-$. From \erefs{phlabel}{khlabel} we can see the meaning of the $z_n$ kinematical variable. 
In the last line, we recall (see Sec.~\ref{s.setup}) that we have defined the partonic momentum fraction $\zetasym \equiv p_{h,H}^+/k_H^+$ in terms of hadron frame components.  
The power counting in the leading region depicted by  \fref{LeadingRegion} is
\begin{align}
p_h = \parzh{\order{Q}, \order{\frac{M_h^2}{Q}}, \T{0}{} } \, , \qquad
k = \parzh{\order{Q}, \order{\frac{M_h^2}{Q}}, \order{\T{\Lambda}{}} } \, , \qquad
q = \parzh{\order{Q}, \order{Q} , \T{0}{} } \, , 
\end{align}
where $\order{\T{\Lambda}{}}$ indicates spatial transverse momenta of the order of the hadronic mass scale $M_h$. The $\zn$ in \eref{zvars} is 
\begin{equation}
\frac{p_{h,H}^+}{q_H^+} = \frac{z}{2} + \sqrt{\frac{z^2}{4} - \frac{M_h^2}{Q^2}} \, .
\end{equation}
The $z_n$ variable has the advantage that it allows us to express the exact momentum components in terms of lightcone plus-momentum fractions.

The lab frame is just the hadron frame rotated until the $z$-axis lies along the direction of the incoming electron momentum. Throughout this paper, whenever momentum components are expressed without subscripts, it should be assumed that they correspond to the lab frame. 

The ``parton frame,'' denoted by $p$ subscripts, is defined to be a frame where $k_p^+ = k_H^+$, $\T{k}{p} = \T{0}{}$, and $\T{p}{h,p} = -\zetasym \T{k}{H}$. The
transformation from hadron to parton frame components is 
\begin{align}
p_h &{}= \parz{p_H^+,\frac{M_h^2 + \zetasym^2 \Tscsq{k}{H}}{2 p_H^+},- \zetasym\T{k}{H}}_p = \parz{\zetasym k_H^+,\frac{M_h^2 + \zetasym^2 \Tscsq{k}{H}}{2 \zetasym k_H^+},- \zetasym\T{k}{H}}_p \, , \\
k &{}= \parz{k_H^+,k_H^- - \frac{\Tscsq{k}{H}}{2 k_H^+},\T{0}{}}_p \, .
\end{align}
The parton frame is useful for discussing definitions of fragmentation functions.

\subsection{Structure functions} 
The usual structure function decomposition is\footnote{For this paper, we will not consider polarization effects, so other structure functions will not be written out explicitly.}
\begin{equation}
W^{\mu \nu}_{\text{SI}} = \parz{-g^{\mu \nu} + \frac{q^\mu q^\nu}{q^2}} F_1(z,Q^2,\left\{p_h\right\}) + \frac{\parz{p_h^\mu - q^\mu \frac{p_h \cdot q}{q^2}}\parz{p_h^\nu - q^\nu \frac{p_h \cdot q}{q^2}}}{p_h \cdot q} F_2(z,Q^2,\left\{p_h\right\}) \, .
\end{equation}
Similarly, the decomposition of the tensor at the partonic level is
\begin{equation}
\label{e.partonstruct}
\widehat{W}^{\mu \nu}_{\text{SI}} = \parz{-g^{\mu \nu} + \frac{q^\mu q^\nu}{q^2}} \widehat{F}_1(\hat{z},Q^2) + \frac{\parz{\hat{k}^\mu - q^\mu \frac{\hat{k} \cdot q}{q^2}}\parz{\hat{k}^\nu - q^\nu \frac{\hat{k} \cdot q}{q^2}}}{\hat{k} \cdot q} \widehat{F}_2(\hat{z},Q^2) \, .
\end{equation}
A convenient way to extract partonic structure functions in calculations is by contracting the projection tensors, 
\begin{align}
\widehat{P}_1^{\mu \nu} &= - \parz{\frac{g^{\mu \nu}}{2} + \frac{2 \hat{k}^\mu \hat{k}^\nu}{Q^2 \hat{z}^2}} \, ,  \label{e.hatP1} \\
\widehat{P}_2^{\mu \nu} &= \frac{g^{\mu \nu}}{\hat{z}} +  \frac{12 \hat{k}^\mu \hat{k}^\nu}{Q^2 \hat{z}^3} \, ,
\label{e.hatP2}
\end{align}
so that
\begin{align}
&\widehat{P}_1^{\mu \nu} \widehat{W}_{\mu \nu \, \text{SI}} = \widehat{F}_1(\hat{z},Q^2)\, , \qquad  \widehat{P}_2^{\mu \nu} \widehat{W}_{\mu \nu \, \text{SI}} = \widehat{F}_2(\hat{z},Q^2) \, .
\end{align}
In terms of structure functions, \eref{fact_formula2} is 
\begin{align}
\label{e.fact_formulaF1}
F_1(z,Q^2,\left\{p_h\right\}) &{}= \int \frac{\diff{\zetasym}{}}{\zetasym^2} \widehat{F}_1(\hat{z},Q^2) \, d(\zetasym,\left\{p_h\right\}) + \text{p.s.} \, , \\
\label{e.fact_formulaF2}
F_2(z,Q^2,\left\{p_h\right\}) &{}= \int \frac{\diff{\zetasym}{}}{\zetasym^3} \widehat{F}_2(\hat{z},Q^2) \, d(\zetasym,\left\{p_h\right\}) + \text{p.s.} \, .
\end{align}
One often integrates over all angles and expresses the cross section as a function of $z$ alone. Changing variables in the lab frame then gives
\begin{equation}
\frac{\diff[3]{\vect{p}_h}}{2 E_{p_h} (2 \pi)^3} = \frac{ Q^2}{8 (2 \pi)^3} z \diff{z} \diff{\Omega } + \text{p.s.} \, , \label{e.diffzomega1}
\end{equation}
where $\diff{\Omega } = \diff{\phi} \diff{\cos\parz{\theta}}$. The ``\text{p.s.}'' here is because the $z \approx \zn$ approximation has been used again. Similarly, 
\begin{equation}
\frac{\diff[3]{\vect{\hat{k}}}}{2 E_{\hat{k}} (2 \pi)^3} = \frac{Q^2}{8 (2 \pi)^3} \hat{z} \diff{\hat{z}} \diff{\Omega} + \text{p.s.} \, . 
\label{e.partontrans}
\end{equation}
Then the factorization formula \eref{fact_formula} becomes\footnote{We recall that in Sec.~\ref{s.crosssection} the hatted $\vect{\hat{k}}$ symbol in \eref{partontrans} has been defined as an approximate partonic momentum defined such that in the hadron frame it has zero transverse momentum, i.e. it is exactly collinear to $p_h$, and that the hatted $\hat{z}$ symbol is defined similarly to $z$ in Eq.~\eqref{e.zvars} but referred to $\vect{\hat{k}}$.}, for the single hadron ($n=1$) case, 
\begin{equation}
\label{e.fact_formulazdef}
\frac{\diff{\sigma}{}}{z \diff{z} \diff{\Omega}} = \int_z^1 \frac{\diff{\zetasym}{}}{\zetasym^2} \parz{\frac{\diff{\hat{\sigma}}{}}{\hat{z} \diff{\hat{z}} \diff{\Omega}}} d(\zetasym,\left\{p_h\right\}) + \text{p.s.} \, .
\end{equation}
Contracting $W^{\mu \nu}_{\text{SI},n=1}$ and $\widehat{W}^{\mu \nu}_{\text{SI},n=1}$ in \erefs{tensordecomp}{tensordecomppartonic} with $L_{\mu \nu}$ gives
\begin{align}
& \frac{\diff{\sigma}}{z \diff{z} \diff{\Omega}} = \frac{ \alpha^2}{2 Q^2} \left[ F_1 + F_2 \frac{z}{4} \sin^2 \theta \right] + \text{p.s.} \, , \label{e.diffzom} \\
& \frac{\diff{\hat{\sigma}}}{\hat{z} \diff{\hat{z}} \diff{\Omega}} = \frac{\alpha^2}{2 Q^2} \left[ \widehat{F}_1 + \widehat{F}_2 \frac{\hat{z}}{4} \sin^2 \theta \right] \, . \label{e.diffzomp}
\end{align}
Integrating \eref{diffzom} over $\diff{\Omega }$ gives  \eref{tensordecomp} in a form that is differential only in $z$,
\begin{equation}
\frac{\diff{\sigma}{}}{\diff{z}{}} = \frac{\pi \alpha^2 z}{Q^2} \left[ 2 F_1(z,Q^2) + \frac{z}{3} F_2(z,Q^2) \right] \, + \text{p.s.} \; . \label{e.dzcrosssection}
\end{equation}
In \eref{dzcrosssection}, the ``$\text{p.s.}$'' term is present because we have approximated $\zn \approx z$. 

From \eref{strucexp} and the cross section in \eref{cross}, we also have for the general semi-inclusive $n$-hadron cross section, 
\begin{equation}
\frac{\diff{\sigma}}{\diff{Y}} = \frac{4 (2 \pi)^3 \alpha^2}{Q^6} L_{\mu \nu} W^{\mu \nu}_{\text{SI}} = \frac{4 (2 \pi)^3 \alpha^2}{Q^4} \left[ F_1(z,Q,\left\{p_h\right\}) + \frac{\zn}{4} F_2(z,Q,\left\{p_h\right\}) \sin^2 \theta \right] \, . \label{e.withstructfuncts}
\end{equation}

\section{Factorization}
\label{s.fact}

The following steps are basically a review of steps like those in chapter 6.1 of Ref.~\cite{Collins:2011qcdbook} and similar treatments, but now specialized to the case of fragmentation functions in $e^+e^-$ annihilation rather than for parton densities in deep inelastic scattering. Our focus is on highlighting the kinematical approximations needed to get factorization and verifying that they are unmodified if the observed hadron is an $n$-hadron. The aim is to expand in powers of $\Lambda/Q$ as $Q \to \infty$ where $\Lambda$ was defined in Sec.~\ref{s.setup} to be any other fixed mass scale like $m$, $M_h$, or an intrinsic transverse momentum. That is, 
\begin{equation}
\Lambda \in \left\{m, M_h, \Tsc{k}{}, \dots \right\} \, .
\end{equation}
In QCD, $\Lambda$ would also include $\Lambda_\text{QCD}$. Also, everywhere below we will use the notation
\begin{equation}
k_2 = q - k 
\end{equation}
to abbreviate expressions even though the $k_2$ momentum is not independent of $q$ and $k$.

Before factorization, a graph with the structure of \fref{LeadingRegion} is 
\begin{equation}
W^{\mu \nu}_{\text{SI}} = \frac{1}{4 \pi} \int \frac{\diff[4]{k}}{(2 \pi)^4} {\rm Tr} \left[J_1(k,\left\{p_h\right\}) \gamma^\mu J_2(k_2) \gamma^\nu \right] \, . \label{e.unfact0}
\end{equation}
For approximations in the hard vertices (which are just zeroth order in our discussion), one uses  $k_2^2 = (q - k)^2 = \order{\Lambda^2}$ to find
\begin{equation}
k_H^+ = \frac{Q^2}{2 q_H^-} + \order{\frac{\Lambda^2}{Q}} = q_H^+ + \order{\frac{\Lambda^2}{Q}} \, ,
\end{equation}
and $\zn = z$ to leading power. Thus, we define a ``hatted'' approximation in the hadron frame, 
\begin{equation}
\hat{k} \equiv \parzh{q_H^+,0,\T{0}{}} \, .
\end{equation}
The power counting for $k$ in the hadron frame is 
\begin{equation}
k = \parzh{\order{Q},\order{\frac{\Lambda^2}{Q}},\order{\T{\Lambda}{}}} \, .
\end{equation}
In addition, 
\begin{equation}
q - k = \parzh{\order{\frac{\Lambda^2}{Q}},q_H^- + \order{\frac{\Lambda^2}{Q}}, \order{\T{\Lambda}{}}} \, ,
\end{equation}
so we define another ``hatted'' approximation for the opposite moving antiquark, 
\begin{equation}
\hat{k}_2 \equiv \parzh{0,q_H^-,\T{0}{}} \, .
\end{equation}
The dominant power law behavior for the components of $k_2$ in the hadron frame is
\begin{equation}
k_2 = \parzh{\order{\frac{\Lambda^2}{Q}},\order{Q},\order{\T{\Lambda}{}}} \, .
\end{equation}
To extract the leading power behavior in the subgraphs 
$J_1(k,\left\{p_h\right\})$ and $J_2(k_2)$, we replace the largest components of $k$ by their approximate values, while the small components must be kept exact. To recall why, consider, for example, the propagator denominator  
$k^2 - m_q^2$ inside $J_1(k,\left\{p_h\right\})$. The power law behavior is
\begin{align}
k^2 - m_q^2 {}= 2 k_H^+ k_H^- - \Tscsq{k}{H} - m_q^2 
{}= 2 \parz{q_H^+ + \order{\frac{\Lambda^2}{Q}} } \underbrace{k_H^-}_{\order{\Lambda^2/Q}} - \underbrace{\parz{\Tscsq{k}{H} + m_q^2}}_{\order{\Lambda^2}} = \underbrace{2 q_H^+ k_H^- - \Tscsq{k}{H} - m_q^2}_{\order{\Lambda^2}} + \order{\frac{\Lambda^4}{Q^2}} \, . \label{e.J1power}
\end{align} 
Approximating $k_H^+$ by $\hat{k}_H^+ = q_H^+$ introduces only power suppressed errors, whereas also neglecting $\Tsc{k}{H}$ or $k_H^-$ would introduce uncontrolled leading power errors. Thus, we define a separate ``tilde'' approximation for $k$,
\begin{equation}
k \to \tilde{k} = \parzh{q_H^+,k_H^-,\T{k}{H}} \, , 
\end{equation}
and use the leading power approximation, 
\begin{equation}
\label{e.J1approx}
J_1(k,\left\{p_h\right\}) \to J_1(\tilde{k},\left\{p_h\right\}) \, .
\end{equation}
The hatted variables are appropriate for describing hard partonic scattering with massless on-shell quarks, while the tilde variables enter the separate collinear subgraphs. The partonic $z$ is indicated with a hat, 
\begin{equation}
\hat{z} = \frac{2 \hat{k}{} \cdot q}{Q^2} = \frac{\zn}{\zetasym} \, .
\end{equation}

To make the analogous approximation on $k_2$ in $J_2(k_2)$, it is convenient to analyze its components in a frame that is slightly boosted from the hadron frame such that $k_2$ has exactly zero transverse momentum, and its minus component equals its minus component in the hadron frame. We will indicate components in this frame with a `*'. Then, 
\begin{equation}
\label{e.k2plus}
k_2 = \parz{q_H^+ - k_H^+ - \frac{\Tscsq{k}{H}}{2 \parz{q_H^- - k_H^-}},q_H^-  - k_H^-,\T{0}{}}^* \, .
\end{equation}
Recalling that $q_H^+ - k_H^+ = \order{\Lambda^2/Q}$, the power law behavior for $k_2$ in `*' frame components is 
\begin{equation}
\label{e.k2pluspowers}
k_2 = \parz{\order{\frac{\Lambda^2}{Q}},\order{Q},\T{0}{}}^* \, .
\end{equation}
Let $r$ be a $k_2$-collinear momentum line internal to the $J_2(k_2)$ subgraph, 
\begin{equation}
\label{e.rpluspowers}
r = \parz{\order{\frac{\Lambda^2}{Q}},\order{Q},\order{\T{\Lambda}{}}}^* \, .
\end{equation}
Then the reasoning for $J_2(k_2)$ analogous to \eref{J1power} is to note that a propagator denominator of the form $\sim (k_2 + r)^2 - m_q^2$ has the power-law behavior, 
\begin{align}
\label{e.J2denom}
(k_2 + r)^2 - m_q^2 = 2 k_2 \cdot r + \left( k_2^2 + r^2 - m_q^2 \right)  
= 2 k_2^{*+} r^{*-} + 2 k_2^{*-} r^{*+} + \order{\Lambda^2} \, .
\end{align}
In \eref{J2denom}, dropping the $k_H^-$ in the plus component of \eref{k2plus} results in dropping an $\order{\Lambda^4/Q^3}$ plus component contribution, whereas the dominant order is  $\order{\Lambda^2/Q}$. Dropping $k_H^-$ in the minus component of \eref{k2plus} results in dropping an $\order{\Lambda^2/Q}$ minus component contribution while the dominant order is an $\order{Q}$ contribution: 
\begin{align}
\parz{k_2 + r}^2 - m_q^2 &= k_2^2 + 2 k_2 \cdot r + r^2 - m_q^2 \no
&= 2 k_2^{*+} \parz{k_2^{*-} + r^{*-}}  + 2 k_2^{*-} r^{*+} + r^2 - m_q^2 \no
&= 2 \parz{q_H^+ - k_H^+ - \frac{\Tscsq{k}{H}}{2 \parz{q_H^- - k_H^-}}} \parz{k_2^{*-} + r^{*-}}  + 2 \parz{q_H^- - k_H^-} r^{*+} + r^2 - m_q^2 \no
&= 2 \parz{\underbrace{q_H^+ - k_H^+ - \frac{\Tscsq{k}{H}}{2 q_H^-}}_{\order{\frac{\Lambda^2}{Q}}} + \order{\frac{\Lambda^4}{Q^3}} } \underbrace{\parz{k_2^{*-} + r^{*-}}}_{\order{Q}}  + 2 \parz{\underbrace{q_H^-}_{\order{Q}} + \order{\frac{\Lambda^2}{Q}}} \underbrace{r^{*+}}_{\order{\frac{\Lambda^2}{Q}}} + \underbrace{r^2 - m_q^2}_{\order{\Lambda^2}} \no
&= 2 \parz{q_H^+ - k_H^+ - \frac{\Tscsq{k}{H}}{2 q_H^-} } \parz{k_2^{*-} + r^{*-}}  + 2 q_H^-  r^{*+} + r^2 - m_q^2 + \order{\frac{\Lambda^4}{Q^2}} \; .
\end{align}
Thus, we define the $k_2$ tilde approximation,
\begin{equation}
k_2 \to \tilde{k}_2 \equiv \parz{q_H^+ - k_H^+ - \frac{\Tscsq{k}{H}}{2 q_H^-},q_H^-,\T{0}{}}^* \, ,
\end{equation}
and the effect of this replacement in, for example, \eref{J2denom} is a subleading error of $\order{\Lambda^4/Q^2}$ with respect to the (leading) power $\order{\Lambda^2}$ of the denominator.
So, we make the leading power approximation
\begin{equation}
\label{e.J2approx}
J_2(k_2) \to J_2(\tilde{k}_2) \, .
\end{equation}
Finally, the leading power Dirac components for $J_1(\tilde{k},\left\{p_h\right\})$ and $J_2(\tilde{k}_2)$ in the hadron frame are, respectively, 
\begin{align}
J_1(\tilde{k},\left\{p_h\right\}) &{}= \frac{1}{4} {\rm Tr} \left[\gamma^+ J_1(\tilde{k},\left\{p_h\right\}) \right] \gamma^- + \text{p.s.} \label{e.J1traceapp} \\
J_2(\tilde{k}_2) &{}= \frac{1}{4} {\rm Tr} \left[\gamma^- J_2(\tilde{k}_2) \right] \gamma^+ + \text{p.s.} \label{e.J2traceapp}  \, .
\end{align}
A change in variables from $k_H^+$ to $\tilde{k}_2^{*+}$ is a linear transformation with unit Jacobian, so 
\begin{equation}
\int \diff[4]{k} (\ldots) = \int \diff{k_H^+} \diff{k_H^-} \diff[2]{\T{k}{H}} (\ldots) = 
\int \diff{\tilde{k}_{2}^{*+}} \diff{k_H^-} \diff[2]{\T{k}{H}} (\ldots) \; . \label{e.inttrans}
\end{equation} 
\begin{figure}[h!]
\centering
\includegraphics[width=8cm]
{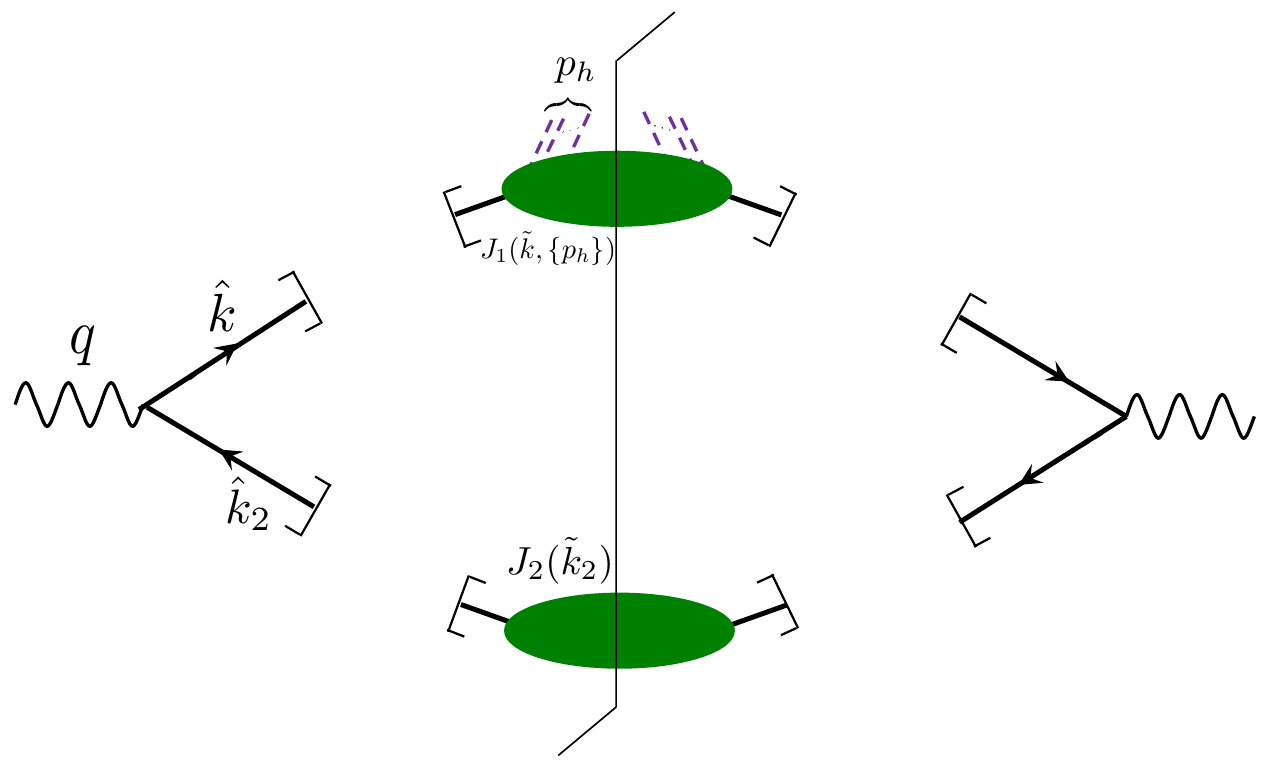}
\caption{The separation into factors in \eref{fact5}. The hooks represent the application of the approximations in \eref{J1approx} and \eref{J2approx}. The top factorized segment corresponds to the first factor in parentheses in \eref{fact5} and the bottom factorized segment corresponds to the second factor in parentheses.}
\label{f.Factorized}
\end{figure}
Making the approximations in \eref{J1approx}, \eref{J2approx}, and \erefs{J1traceapp}{J2traceapp}, and using the variable transformation in \eref{inttrans} allows the hadronic tensor in \eref{unfact0} to be rewritten as 
\begin{align}
&{}W^{\mu \nu}_{\text{SI}} = \frac{1}{16 \pi} {\rm Tr} \left[ \gamma^- \gamma^\mu \gamma^+ \gamma^\nu \right] \int \frac{\diff{\tilde{k}_{2}^{*+}} \diff{k_H^-} \diff[2]{\T{k}{H}}}{(2 \pi)^4}  \parz{\frac{1}{2} {\rm Tr} \left[\gamma^+ J_1(\tilde{k},\left\{p_h\right\}) \right]} \parz{\frac{1}{2} {\rm Tr} \left[\gamma^- J_2(\tilde{k}_2) \right]}  + \text{p.s.} \, .
\label{e.fact4}
\end{align}
The factor in the first set of parentheses is independent of $k_2^{*+}$ while the factor in the second set of parentheses is independent of $k_H^-$ and $\T{k}{H}$. Thus, rearranging the placement of integral symbols in \eref{fact4} gives 
\begin{align}
&{}W^{\mu \nu}_{\text{SI}} =\frac{1}{4 Q^2} {\rm Tr} \left[ \sla{\hat{k}} \gamma^\mu \sla{\hat{k}}_2 \gamma^\nu \right] \parz{ \int \frac{\diff{k_H^-} \diff[2]{\T{k}{H}}}{(2 \pi)^4} {\rm Tr} \left[\frac{\gamma^+}{2} J_1(\tilde{k},\left\{p_h\right\}) \right]} \parz{ \int \frac{\diff{\tilde{k}_2^{*+}}}{2 \pi} {\rm Tr} \left[\frac{\gamma^-}{2} J_2(\tilde{k}_2) \right]}  + \text{p.s.}\, ,
\label{e.fact5}
\end{align}
where we have also multiplied and divided by $\hat{k}_H^+ \hat{k}_{2H}^- = Q^2/2$ to put the overall factor into a more recognizable form. The separation into factors in \eref{fact5} is shown diagrammatically in \fref{Factorized}. 

Up to standard issues related to renormalization and ultraviolet divergences, the last factor in parenthesis is 1 by unitarity. By very general arguments (see sections 12.5-12.7 of Ref.~\cite{Collins:2011qcdbook}), the complete sum and integral over final states in $J_2(\tilde{k}_2)$ is insensitive to infrared  contributions. Here this can be demonstrated by calculating its lowest order massless perturbative approximation. The spin-summed lowest order subgraph is $J_2(\tilde{k}_2) = \tilde{\sla{k}}_2 (2 \pi) \delta_+(\tilde{k}_2^2)$, so
\begin{equation}
\int \frac{\diff{\tilde{k}_2^{*+}}}{2 \pi} {\rm Tr} \left[\frac{\gamma^-}{2} J_2(\tilde{k}_2) \right] \stackrel{\text{L.O.}}{\to} \int \diff{\tilde{k}_2^{*+}} {\rm Tr} \left[\frac{\gamma^-}{2} \sla{\tilde{k}}_2 \right] \delta_+(\tilde{k}_2^2) = 1 \, .
\end{equation}
Then, 
\begin{align}
W^{\mu \nu}_{\text{SI}} 
&{}=\int_{\zn}^1 \diff{\zetasym} \frac{1}{2 Q^2} \delta(1-\zn/\zetasym) {\rm Tr} \left[ \sla{\hat{k}} \gamma^\mu \sla{\hat{k}}_2 \gamma^\nu \right] \parz{ \int \frac{\diff{k_H^-} \diff[2]{\T{k}{H}}}{2 \zetasym (2 \pi)^4}  {\rm Tr} \left[\frac{\gamma^+}{2} J_1(\tilde{k},\left\{p_h \right\}) \right]} + \text{p.s.} \no  
&{}=\int_{\zn}^1 \frac{\diff{\zetasym}}{\zetasym^2} \parz{ \frac{1}{2 Q^2} \delta(1-\zn/\zetasym) {\rm Tr} \left[ \sla{\hat{k}} \gamma^\mu \sla{\hat{k}}_2 \gamma^\nu \right]} \parz{ \zetasym^2 \int \frac{\diff{k_H^-} \diff[2]{\T{k}{H}}}{2 \zetasym (2 \pi)^4} {\rm Tr} \left[\frac{\gamma^+}{2} J_1(\tilde{k},\left\{p_h \right\}) \right]} + \text{p.s.} \, .
\label{e.fact6}
\end{align}
Here, we have also replaced $\zn$ in the expressions for $\tilde{k}^+$ and $\hat{k}^+$ by a generic momentum fraction $\zetasym$, which we then set to $\zetasym = \zn$ by inserting a $\delta$-function. The lower limit starts from $z_n$ because in higher orders $q_H^+ \geq k_H^+$ in $\hat{z}=\zn/\xi = k_H^+/q_H^+$. 
\begin{figure}[h!]
\centering
\includegraphics[width=6cm]
{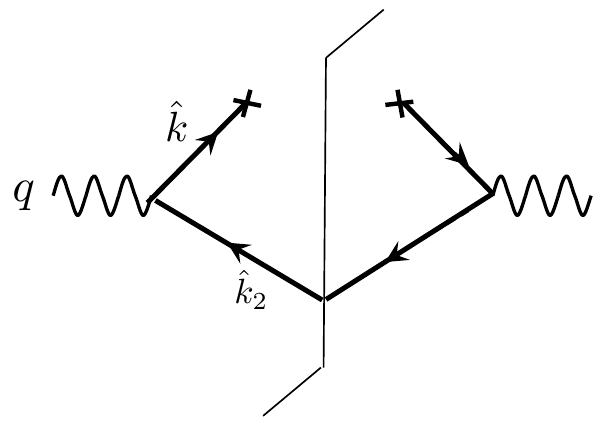}
\caption{The zeroth order contribution to $\widehat{W}_\text{SI}^{\mu \nu}(\hat{z})$, the hadronic tensor  of \eref{WSIhat} for semi-inclusive production of an on-shell massless parton.}
\label{f.ZerothOrder}
\end{figure}

To identify the final factorized structure, it is necessary to compare \eref{fact6} with the zeroth order contribution to the partonic $\widehat{W}^{\mu \nu}_{\text{SI}}(\hat{z})$ from \eref{WSIhat} and shown graphically in \fref{ZerothOrder}. It is
\begin{align}
\widehat{W}_\text{SI}^{\mu \nu}(\hat{z}) &{}= \frac{1}{4 \pi} \int \frac{\diff[4]{\hat{k}_2}}{(2\pi)^4} {\rm Tr} \left[ \sla{\hat{k}} \gamma^\mu \sla{\hat{k}}_2 \gamma^\nu \right] (2 \pi)^4 \delta^{(4)}(q - \hat{k} - \hat{k}_2) (2 \pi) \delta_+ (\hat{k}_2^2) = \frac{1}{2 Q^2} {\rm Tr} \left[ \sla{\hat{k}} \gamma^\mu \sla{\hat{k}}_2 \gamma^\nu \right] \delta(1 - \zn/\zetasym) \, , \label{e.hardW}
\end{align}
with $\hat{k}_2 = q - \hat{k}$ after the integral on the $\hat{k}_2$ line.
Then \eref{fact6} becomes the final factorization formula,
\begin{align}
W^{\mu \nu}_{\text{SI}} 
&{}=\int_{\zn}^1 \frac{\diff{\zetasym}}{\zetasym^2} \,  \widehat{W}_\text{SI}^{\mu \nu}(\zn/\zetasym) \, d(\zetasym,\left\{p_h\right\}) + \text{p.s.} \, ,
\label{e.fact7}
\end{align}
if we are able (see \sref{ffs} below) to make the identification for the collinear fragmentation function
\begin{equation}
d(\zetasym,\left\{p_h\right\}) = \zetasym^2 \int \frac{\diff{k_H^-} \diff[2]{\T{k}{H}}}{2 \zetasym (2 \pi)^4} {\rm Tr} \left[\frac{\gamma^+}{2} J_1(\tilde{k},\left\{p_h \right\}) \right] \, . \label{e.ffcandidate}
\end{equation}
Contracting \eref{fact7} with $L^{\mu \nu}$ produces \eref{fact_formula}. Using the projectors of Eqs.~\eqref{e.hatP1},\eqref{e.hatP2} on Eq.~\eqref{e.hardW}, we get 
\begin{equation}
\widehat{F}_1 = \frac{z_n}{\xi} \, \delta \left( 1 - \frac{z_n}{\xi} \right) \; , \quad \widehat{F}_2 = -2 \, \delta \left( 1 - \frac{z_n}{\xi} \right) \; , 
\label{e.hatF1F2}
\end{equation}
from which~\erefs{fact_formulaF1}{fact_formulaF2} read (after making the $z \approx \zn$ leading-$\Lambda/Q$ approximation) 
\begin{equation}
F_1 {}= \frac{1}{z} d(z,\left\{p_h\right\}) \; , \quad 
F_2 {}= -\frac{2}{z^2} d(z,\left\{p_h\right\}) \, . \label{e.F2part}
\end{equation}
In going from \eref{fact5} to the first line of \eref{fact6} it was necessary to factor a $\zetasym$ into the hard part and a factor of $1/\zetasym$ into the part in parentheses (which will become the fragmentation function) in order to get the hard $\widehat{W}_\text{SI}^{\mu \nu}(\hat{z})$. This ultimately makes the fragmentation function a density in the parton frame phase space $\diff{\zetasym} \diff[2]{\T{p}{hp}}$, as seen in \sref{ffs} below. The $\zetasym$ factored into the hard part ensures that it has the final state phase space of the partonic cross section. The trend continues to higher orders and the steps above confirm that they are independent of the identity of the observed final state $\left\{ p_h \right\}$. In particular, they are independent of whether it is a single hadron or an $n$-hadron. 
The Lorentz invariant phase space associated with the observed $n$-hadron in the $J_1(\tilde{k},\left\{p_h\right\})$ subgraph is still the initial one in \eref{basicps} from the full cross section expression, 
\begin{equation}
\label{e.basicps2}
\frac{\diff{}}{\diff{Y}}  = 2 E_{h_1} (2 \pi)^3 \frac{\diff{}}{\diff[3]{\vect{p}_{h_1}}}
2 E_{h_2} (2 \pi)^3 \frac{\diff{}}{\diff[3]{\vect{p}_{h_2}}}
\times \cdots \times 2 E_{h_n} (2 \pi)^3 \frac{\diff{}}{\diff[3]{\vect{p}_{h_n}}} \, .
\end{equation}
Even though the partonic transverse momentum has been integrated out in \eref{fact6} the cross section and $W^{\mu \nu}_{\text{SI}}$ are still differential in the hadronic transverse momentum. 
 
To summarize this section, nothing in the steps above leading to the factorization formula in \eref{fact6} and the fragmentation function in \eref{ffcandidate} distinguishes between the single hadron and $n$-hadron situations.

\section{Fragmentation functions}
\label{s.ffs}

\subsection{Definition}

To complete the discussion of factorization, it is still necessary to confirm that \eref{ffcandidate} matches the usual operator definition of a fragmentation function, so in this section we review the steps for defining fragmentation functions while generalizing the treatment to the $n$-hadron case. The following is a slight variation of the discussion in section 12.4 of Ref.~\cite{Collins:2011qcdbook}, based on Ref.~\cite{Collins:1981uw}, now tailored to incorporate multihadron fragmentation functions. For most of this discussion, we will drop explicit spin indices to simplify notation, but they should be understood to be implicit where necessary.

The basic definition (still working in the non-gauge theory) of a bare transverse momentum dependent single hadron fragmentation function is
\begin{align}
2 \zetasym (2 \pi)^3 d(\zetasym,-\zetasym \T{k}{H},p_h) \langle \vect{k}_1 | \vect{k}_2 \rangle &{}= \langle \vect{k}_1 | \vect{k}_2 \rangle  \left. \frac{\diff{N}}{\diff{Y}{}} \right|_{\text{single hadron}} = 
 \SumInt_X \langle \vect{k}_1 |p_h, X \rangle \langle p_h, X | \vect{k}_2 \rangle \, , \label{e.basicdef}
\end{align}
or, 
\begin{align}
d\parz{\zetasym,-\zetasym \T{k}{H},p_h} \langle \vect{k}_1 | \vect{k}_2 \rangle = \frac{1}{2 \zetasym (2 \pi)^3} \SumInt_X \langle \vect{k}_1 |p_h, X \rangle \langle p_h, X | \vect{k}_2 \rangle \, . \label{e.basicdef3}
\end{align}
The reason for expressing the parton momentum arguments in the variables above is that it simplifies discussions of fragmentation functions in the parton frame where the parton transverse momentum is fixed to zero and all transverse momentum is carried by the hadron transverse momentum, $\T{p}{hp} = -\zetasym \T{k}{H}$.
The $|\vect{k}_1\rangle$ and $|\vect{k}_2\rangle$ are states created by the action of lightcone creation and annihilation operators $b_{\vect{k}}^\dagger$ and $b_{\vect{k}}$, which satisfy lightcone anticommutation relations, acting on the vacuum, and to keep notation compact we have suppressed other quantum numbers\footnote{These operators should not be confused with  the equal time creation operators $a_{\vect{p}}^\dagger$ used for the on-shell hadronic asymptotic ``out'' states in \eref{hadtensSI}.}. Since we are working with non-normalizable quark momentum eigenstates we start with off-diagonal states $| \vect{k}_1 \rangle$ and $| \vect{k}_2 \rangle$. The reason for the $2 \zetasym (2 \pi)^3$ on the left-hand side of \eref{basicdef} will be made clear below. 

To generalize \eref{basicdef} to the $n$-hadron case, one defines
\begin{align}
2 \zetasym (2 \pi)^3 d(\zetasym,-\zetasym \T{k}{H},\left\{ p_h \right\}) \langle \vect{k}_1 | \vect{k}_2 \rangle &{}= \langle \vect{k}_1 | \vect{k}_2 \rangle \frac{\diff{N}}{\diff{Y}{}}  \no
&{}= \langle \vect{k}_1 | \vect{k}_2 \rangle 2 E_{\hv{1}} (2 \pi)^3 \frac{\diff{}}{\diff[3]{\vect{p}_{h_1}}} 2 E_{\hv{2}} (2 \pi)^3 \frac{\diff{}}{\diff[3]{\vect{p}_{h_2}}} \times \cdots \times 2 E_{\hv{n}} (2 \pi)^3 \frac{\diff{}}{\diff[3]{\vect{p}_{h_n}}} N \no 
&{} = \SumInt_X \langle \vect{k}_1 |\hv{1},\hv{2},\dots,\hv{n}, X \rangle \langle \hv{1},\hv{2},\dots,\hv{n}, X | \vect{k}_2 \rangle \, , \label{e.nhadrondef}
\end{align}
or, 
\begin{align}
d\parz{\zetasym,-\zetasym \T{k}{H},\left\{ p_h \right\}} \langle \vect{k}_1 | \vect{k}_2 \rangle = \frac{1}{2 \zetasym (2 \pi)^3} \SumInt_X \langle \vect{k}_1 |\hv{1},\hv{2},\dots,\hv{n}, X \rangle \langle \hv{1},\hv{2},\dots,\hv{n}, X | \vect{k}_2 \rangle \, . \label{e.nhadron3}
\end{align}
The $\diff{Y}{}$ has just been generalized to a multihadron phase space. 
Using the expressions for $b_{\vect{k}}^\dagger$ and $b_{\vect{k}}$ in terms of the quark field operator $\psi(x)$ (and retracing the steps in Sec.~(6.7.3) of Ref.~\cite{Collins:2011qcdbook}) puts the definition in the more familiar form (see Eq.~(12.39) of Ref.~\cite{Collins:2011qcdbook}) with coordinate space field operators, but now with a system of $n$ observed hadrons in the final state with $n$ not necessarily equal to 1, 
\begin{align}
& d\parz{\zetasym,-\zetasym \T{k}{H},\left\{ p_h \right\}} \, \no
& \; \equiv {\rm Tr} \frac{1}{4 \zetasym} \SumInt_X \int \frac{\diff{x^-}{} \diff[2]{\T{x}{}}{}}{(2 \pi)^3} e^{i x^- p_{h,H}^+/\zetasym - i \T{x}{} \cdot \T{k}{H} } \langle 0 | \gamma^+ \psi(x/2) |\hv{1},\hv{2},\dots,\hv{n}, X \rangle \langle \hv{1},\hv{2},\dots,\hv{n}, X |\overline{\psi}(-x/2)| 0 \rangle \, ,
\label{e.ffdef}
\end{align}
with $x = (0,x^-,\T{x}{})$ and where we have kept only the trace of $\gamma^+$ corresponding to the unpolarized fragmentation function. Translated into Feynman rules, this TMD fragmentation function is
\begin{equation}
d(\zetasym,-\zetasym \T{k}{H},\left\{ p_h \right\}) = \frac{1}{4 \zetasym} \int \frac{\diff{k^-_H}}{(2 \pi)^4} {\rm Tr} \left[\gamma^+  \picineq{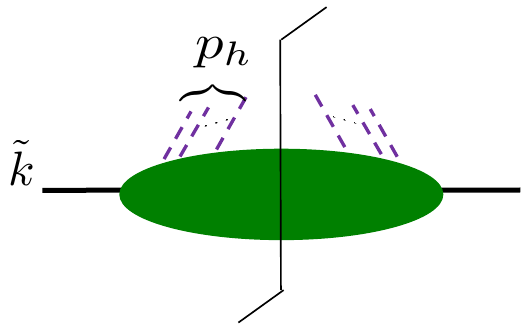} \right]_{k_H^+ = p_{h,H}^+/\zetasym} \, .
\label{e.tmdgraphical}
\end{equation}
The cut final state in green is the usual sum of Feynman diagrams integrated over the final state phase space.
The collinear version is obtained by integrating over the
parton transverse momentum in the form of $-\zetasym \T{k}{H}$,
\begin{align}
d\parz{\zetasym,\left\{ p_h \right\};\mu} &\equiv \zetasym^2 \int^{\sim \mu^2} \diff[2]{\T{k}{H}} d\parz{\zetasym,-\zetasym \T{k}{H},\left\{ p_h \right\}} \no 
&= \frac{\zetasym}{4} \int^{\sim \mu^2} \frac{\diff{k^-_H} \diff[2]{\T{k}{H}}}{(2 \pi)^4} {\rm Tr} \left[\gamma^+  \picineq{FFfeyn.pdf} \right]_{k_H^+ = p_{h,H}^+/\zetasym} \, .
\label{e.collinearoperator}
\end{align}
This recovers the second factor in parentheses on the second line of \eref{fact6}, and it confirms that the operator definition matches what we found in \eref{ffcandidate} for the factorized expression. 

In \eref{collinearoperator}, we have inserted a $\sim \mu^2$ on the integral as a reminder of the need to regulate the ultraviolet divergent transverse momentum integral (not necessarily with a cutoff) around a scale $\mu^2$, and we have restored an explicit auxiliary $\mu$ argument. As in the previous section, nothing changes relative to the treatment of the single hadron case except that the number density is now multidifferential in the $\left\{ p_h \right\}$ momenta and the observed final state now includes $n$ hadrons. 

Up to a careful treatment of renormalization, the definition in \eref{nhadron3} is the standard fragmentation function definition that appears in the factorization in \eref{fact6} and \eref{fact7}.  When $n=1$, the overall $2 \zetasym (2 \pi)^3$ on the left-hand side of \eref{nhadrondef} ensures that the $d\parz{\zetasym,-\zetasym \T{k}{H},\left\{ p_h \right\}}$ has the interpretation of a number density, with respect to $\diff{\zetasym}{} \diff[2]{\T{p}{hp}}{}$, of hadrons associated with the fragmenting quark. In gauge theories a Wilson line needs to be accounted for as well. 

The above definition does not specify anything with regard to the internal relative momenta of the hadrons that make up the $n$-hadron system. 
In applications, one often integrates over some subset of these momentum components. 
We will call a function that results from such a simplifying integration a ``reduced'' fragmentation function. Different reduced fragmentation functions will result from different choices of variables and different ranges of integration, but the way that one connects them to fundamental operator matrix elements ultimately depends only on the variable convention.

In applications to cross sections with $n>1$, it is often convenient to change variables and write 
\begin{equation}
\frac{\diff{}}{\diff{Y}{}} \cdots = \parz{ 2 E_{p_h} (2 \pi)^3 \frac{\diff{}{}}{\diff[3]{\vect{p}_h}{}}} \parz{ \frac{\diff{}{}}{\diff{Y'}}}_H \cdots \label{e.Yprime}
\end{equation}
where $\diff{Y'}{}$ is the phase space of the relative internal momenta of hadrons making up the $n$-hadron system. By expressing it in hadron frame components, one may eliminate the primed variables' dependence on $p_h$. Then the set of factors and derivatives in the first set of parentheses is the usual Lorentz invariant phase space for the total momentum of the $n$-hadron system. A particular reduced multihadron fragmentation function is obtained when one integrates over some choice of subset of the momenta that make up $Y'$.
The integral
\begin{equation}
\int_{V} \diff{Y'}{} d\parz{\zetasym,-\zetasym \T{k}{H},\left\{ p_h \right\}} 
\end{equation}
is then a density, again with respect to $\diff{\zetasym}{} \diff[2]{\T{p}{hp}}{}$, of $n$-hadron systems whose relative internal momenta lie in an integration volume $V$. Recall that the integration volume $V$ must be restricted to $M_h^2 \ll Q^2$ for factorization to hold.  

A potential point of confusion with the notation is that it might appear that the integral over $\T{p}{hp}$ that reduces the TMD to the collinear fragmentation function corresponds to an integral over a physically measured external momentum in a full cross section. That is not the case, however, because the parton frame is not fixed. It changes relative to the center of mass of the physical process as the internal parton momentum is integrated over. The cross section itself remains differential in the physically observed transverse momentum of the final state hadrons, even after the integrals over $\T{p}{hp}$ have been carried out.  

Note that the factor of $\zetasym$ on the left-hand side of 
\eref{nhadrondef} is necessary for the definition of the fragmentation function to 
match the factorization result in \eref{fact6}. Since the fragmentation function definitions are independent of the number of hadrons in the external state, then so are the evolution equations.

\subsection{Interpretations}
\label{s.interps}

As with any typical relativistic interacting quantum field theory that requires renormalization, one cannot define an exactly literal probability density for finding a certain number of elementary particles inside an asymptotic state. At least some properties needed for a literal probability density are violated. It is more precise, therefore, to speak of quasiprobability distributions in the context of partonic correlation functions, in the sense that for most (but not all) practical situations it is possible to treat objects like fragmentation functions \emph{as if} they are  probability distributions. In particular, one may use them to construct expectation values, and the structure of factorization formulas like \eref{fact_formula} matches the probabilistic parton model intuition wherein an elementary partonic cross section is convoluted with a number density for producing an $n$-hadron state in a given differential region of phase space, in a way that closely follows probabilistic intuition. Furthermore, the definition of the fragmentation function in \eref{ffdef} has a natural probabilistic interpretation when viewed as an extension of the kind of probability distributions that do arise in theories that are ultraviolet finite. Provided that the usual caveats associated with renormalization, etc., are kept in mind, it can be understood as a number density, with respect to a parton momentum fraction $\xi$, of objects carrying momentum $p_h$ in a single fragmenting parton. The same interpretation continues to apply if the object is an $n$-hadron 
(in a restricted region of $\diff{Y'}{}$) 
rather than just a single isolated hadron. 

Therefore, we disagree with the assertion in Ref.~\cite{Pitonyak:2023gjx} that the standard fragmentation function definition lacks an acceptable number density interpretation within a parton model framework when it is extended to the multihadron case, at least so long as the usual limitations of number density interpretations alluded to above, and that apply to most types of QCD correlation functions,
are acknowledged. 
When we discuss parton model interpretations below, we will assume that all such caveats are understood and taken into account.

The set of steps leading to \eref{fact_formula}, with the standard $n$-hadron fragmentation function definition appearing in the factorization formula, are the main results of this paper. In the remaining sections, we will focus on showing the details of variable transformations needed to rewrite cross sections in forms more familiar from phenomenological applications in various choices of variables for the external hadrons. 

As mentioned in the previous subsection, the internal kinematical details of the $n$-hadron state are often unknown, and so it is usually convenient to define a kind of ``reduced'' fragmentation function by integrating 
over some subset of the variables that make up $\diff{Y'}{}$.  
The choice of variables used to do this is arbitrary, and it does not impact the basic definition above or the validity of its interpretation. 

Using Eq.~\eqref{e.hatF1F2}, a semi-inclusive cross section for production of an $n$-hadron at zeroth order is 
\begin{align}
\label{e.nfact_formula_zeroth}
\frac{\diff{\sigma}{}}{\diff{Y}} &= \int_z^1 \frac{\diff{\zetasym}{}}{\zetasym^2} \parz{2 E_{\hat{k}} (2 \pi)^3 \frac{\diff{\hat{\sigma}}{}}{\diff[3]{\vect{\hat{k}}}{}} } d(\zetasym,\left\{p_h\right\}) + \text{p.s.} = \frac{8 (2 \pi)^3}{Q^2} \int_z^1 \frac{\diff{\zetasym}{}}{\zetasym^2} \parz{ \frac{\diff{\hat{\sigma}}{}}{\hat{z} \diff{\hat{z}} \diff{\Omega}} } d(\zetasym,\left\{p_h\right\}) + \text{p.s.} \no
 &= \frac{8 (2 \pi)^3}{Q^2} \int_z^1 \frac{\diff{\zetasym}{}}{\zetasym^2} \parz{ \frac{\alpha^2}{2 Q^2} \left[ \widehat{F}_1 + \widehat{F}_2 \frac{\hat{z}}{4} \sin^2 \theta \right] } d(\zetasym,\left\{p_h\right\}) + \text{p.s.} + \order{\alpha_s} \no
  &= \frac{4 \alpha^2 (2 \pi)^3}{z Q^4} \parz{  1 -  \frac{1}{2} \sin^2 \theta } d( z,\left\{p_h\right\}) + \text{p.s.} + \order{\alpha_s} \, .
\end{align}
Recalling 
\begin{equation}
\frac{\diff{\sigma}}{\diff{Y}} = 2 E_{p_h} (2 \pi)^3 \frac{\diff{\sigma}{}}{\diff[3]{\vect{p}_h}{} \diff{Y'}} = \frac{8 (2 \pi)^3}{ Q^2} \frac{\diff{\sigma}{}}{ z \diff{z} \diff{\Omega } \diff{Y'}} \, 
\end{equation}
gives
\begin{equation}
\label{e.generaldz}
\frac{\diff{\sigma}{}}{ \diff{z} \diff{\Omega } \diff{Y'}} = \frac{\alpha^2}{4 Q^2} \parz{1 + \cos^2 \theta} d(z,\left\{p_h\right\}) + \text{p.s.} + \order{\alpha_s} \, ,
\end{equation}
which matches Eq.(12.29) of Ref.~\cite{Collins:2011qcdbook} after the change in variables in \eref{diffzomega1}. The only difference is the presence of the extra phase space $\diff{Y'}{}$.
Writing a more specific version of a parton model expression for $n>1$ amounts to choosing a set of variables for the $Y'$ phase space. 

\section{Single hadron inclusive production}
\label{s.singlehad}

Now we review the variable transformations that recover the familiar angular-integrated, $z$-dependent factorization formula for the single hadron production case. 
When specializing to $n=1$, \eref{fact7} becomes
\begin{align}
W^{\mu \nu}_{\text{SI}} 
&{}=\int_{z}^1 \frac{\diff{\zetasym}}{\zetasym^2} \,  \widehat{W}_\text{SI}^{\mu \nu}(z/\zetasym) \, d(\zetasym,p_h) + \text{p.s.} + \order{\alpha_s} \, .
\label{e.factsingle}
\end{align}
Here we have replaced $z_n$ everywhere with $z$ to obtain a completely leading power formula. We have also included an ``$\order{\alpha_s}$'' to indicate that we are working with the zeroth order $\widehat{W}_\text{SI}^{\mu \nu}(z/\zetasym)$.
Substituting \eref{F2part} into \eref{dzcrosssection} for the special case of $n = 1$ gives 
\begin{align}
\frac{\diff{\sigma}{}}{\diff{z}{}} = \frac{4 \pi \alpha^2}{3 Q^2} d(z,p_h) \, + \text{p.s.} + \order{\alpha_s}  = \sigma_0 d(z,p_h) \, + \text{p.s.} + \order{\alpha_s}  \, , \label{e.dzcrosssectionn1}
\end{align}
which, after power suppressed errors are dropped, is the standard zeroth order expression. Another way to write it is 
\begin{align}
\frac{1}{\sigma_0} \frac{\diff{\sigma}{}}{\diff{z}{}}  = d(z,p_h) \, + \order{\frac{\Lambda^2}{z Q^2}} + \order{\alpha_s}  \, . \label{e.dzcrosssectionn2}
\end{align}
Here
the power suppressed error term is written out with the typical $1/z$ dependence shown explicitly to emphasize the need to avoid the $z \to 0$ region to maintain factorization. Given the form of \eref{dzcrosssectionn2}, it might be tempting to drop the error terms and use the definition of the inclusive cross section to write
\begin{align}
\sum_{\substack{\text{hadron} \\ \text{types}}} \int_0^1 \diff{z} \frac{1}{\sigma_0} \frac{\diff{\sigma}{}}{\diff{z}{}} = \sum_{\substack{\text{hadron} \\ \text{types}}} \int_0^1 \diff{z}  d(z,p_h) = \langle N \rangle \, \label{e.dzcrosssectionn3}
\end{align}
with $\langle N \rangle$ being a total hadron multiplicity. The sum over hadron types only includes one term since there is only one hadron flavor in our treatment. But the error terms in 
\eref{dzcrosssectionn2} diverge in the integral over $z$ down to $0$. At a minimum, a lower bound must be imposed on $z$ such that $z \gg \Lambda^2 /Q^2$. 

Thus, to a rough approximation, one might expect to find a relation of the form
\begin{align}
\sum_{\substack{\text{hadron} \\ \text{types}}}
\int_{z_\text{min}}^1 \diff{z}  d(z,p_h;\mu) \approx \langle N \rangle (z_\text{min},\mu) \, \label{e.partonmodel}
\end{align}
where $z_\text{min}$ is a cutoff that is much larger than $\order{\Lambda^2/Q^2}$ and we have restored the explicit dependence on a scale $\mu$ in the arguments to emphasize the presence of a scale ambiguity on the right-hand side. Given the discussion in the previous section, the fragmentation function is still the hadron density, now differential in $\diff{z}$. But the dependence of $\langle N \rangle$ on auxiliary variables $\mu$ and $z_\text{min}$ highlights the limitations of a completely straightforward and unambiguous interpretation outside of a parton-model-like approximation. 

Another way to obtain the above is by substituting Eq.~\eqref{e.hatF1F2} into \eref{diffzomp} and \eref{fact_formulazdef} which gives
\begin{align}
\frac{\diff{\sigma}}{\diff{z}} &{}= z \int \diff{\Omega } \int_z^1 \frac{\diff{\zetasym}{}}{\zetasym^2} \parz{\frac{\diff{\hat{\sigma}}{}}{\hat{z} \diff{\hat{z}} \diff{\Omega}}} d(\zetasym,p_h) + \text{p.s.} = z \int_z^1 \frac{\diff{\zetasym}{}}{\zetasym^2} \parz{\frac{\diff{\hat{\sigma}}{}}{\hat{z} \diff{\hat{z}}}} d(\zetasym,p_h) + \text{p.s.}  \no
&{}= z \int_z^1 \frac{\diff{\zetasym}{}}{\zetasym^2} \parz{\frac{z \alpha^2 \pi}{Q^2} \frac{4}{3} \delta\parz{\zetasym - z}} d(\zetasym,p_h) + \text{p.s.} + \order{\alpha_s} = \frac{4 \pi \alpha^2}{3 Q^2} d(z,p_h) + \text{p.s.} + \order{\alpha_s} \, . \label{e.diffzsimp}
\end{align}
The zeroth order partonic cross section 
follows by a direct calculation of the zeroth order Feynman graph in \fref{ZerothOrder}. One more direct way to arrive at \eref{diffzsimp} is to simply calculate \eref{generaldz} with $\diff{Y'} \to 1$ and integrate over $\diff{\Omega}$.

\section{Dihadron inclusive production}
\label{s.dihadronexample}

Next, we review the steps for changing variables to reduce the cross section expression for dihadron production to other familiar forms. To simplify the discussion we will continue to assume that each hadron in a dihadron pair has the same mass $m$. The minimum invariant mass $M_h$ is then $2m$.

For $n=2$, the dihadron cross section in \eref{fact_formula} becomes
\begin{equation}
\label{e.fact_formuladi}
2 E_{\hv{1}} (2 \pi)^3 2 E_{\hv{2}} (2 \pi)^3 \frac{\diff{\sigma}{}}{\diff[3]{\vect{p}_{h_1}}{} \diff[3]{\vect{p}_{h_2}}{}} = \int_z^1 \frac{\diff{\zetasym}{}}{\zetasym^2} \parz{2 E_{\hat{k}} (2 \pi)^3 \frac{\diff{\hat{\sigma}}{}}{\diff[3]{\vect{\hat{k}}}{}} } d(\zetasym,\left\{ p_h \right\}) + \text{p.s.} \, ,
\end{equation}
with the fragmentation function defined as in \eref{collinearoperator}. For factorization to hold, the kinematics need to be restricted to $M_h^2/z \ll Q^2$. 

A standard way to characterize the dihadron momentum that is common to many treatments is with the variables
\begin{equation}
p_h \equiv \hv{1} + \hv{2} \, , \qquad R \equiv \frac{\hv{1} - \hv{2}}{2} \, . \label{e.variablesa}
\end{equation}
We define
\begin{align}
{\zn}_1 \equiv \frac{p_{h_1,H}^+}{q^+_H}  \, , \qquad {\zn}_2 \equiv \frac{p_{h_2,H}^+}{q^+_H} \, , \qquad
z_1 \equiv \frac{2 p_{h_1} \cdot q}{Q^2}  \, , \qquad z_2 \equiv \frac{2 p_{h_2} \cdot q}{Q^2} \, ,
\end{align}
where $\zn = {\zn}_1 + {\zn}_2$. We also define partonic momentum fractions analogous to $\zetasym$, 
\begin{equation}
\zetasym_1 \equiv \frac{p_{h_1,H}^+}{k_H^+} = \zetasym \frac{p_{h_1,H}^+}{p_{h,H}^+} \, , \qquad \zetasym_2 \equiv \frac{p_{h_2,H}^+}{k_H^+} = \zetasym \frac{p_{h_2,H}^+}{p_{h,H}^+} \, ,
\end{equation}
with $\zetasym = \zetasym_1 + \zetasym_2$. It is also useful to define the purely external kinematical variables, 
\begin{equation}
x_1 \equiv \frac{p_{h_1,H}^+}{p_{h,H}^+} = \frac{{\zn}_1}{\zn} \, , \qquad x_2 \equiv \frac{p_{h_2,H}^+}{p_{h,H}^+} = \frac{{\zn}_2}{\zn} \, , \qquad \text{so} \qquad \zetasym_1 \zetasym_2 = \zetasym^2 x_1 x_2 \, . \label{e.xdefs}
\end{equation}
In the zeroth order parton model approximation, $\zetasym \to \zn \to z$ after the integral over $\zetasym$, in which case $\zetasym_1 \to z_1$ and $\zetasym_2 \to z_2$. 
We take the standard definition of the parameter $\zeta$  that quantifies the asymmetry in the hadrons' relative momentum, 
\begin{equation}
\zeta \equiv \frac{p_{h_1,p}^+ - p_{h_2,p}^+}{p_{h,p}^+} = \frac{{\zn}_1 - {\zn}_2}{\zn} = \frac{z_1 - z_2}{z} + \text{p.s.} \, .
\end{equation}

One may integrate over some subset of kinematical variables describing the relative momentum of the hadrons to get a simpler form. The exact form depends on the choice of variables used to describe the phase space of $\vect{p}_{h_1}$ and $\vect{p}_{h_2}$.  We illustrate this with two examples below.

\subsection{Variable choice A}
\label{s.varA}

One standard choice is to use the variables above directly and obtain a cross section that is differential in $M_h$ and $z$. 
In the hadron frame, 
\begin{align}
p_{h_1} = \parz{\frac{\zn (1+\zeta) Q}{2 \sqrt{2}}, \frac{\parz{m^2 +\Tscsq{R}{H}}\sqrt{2}}{\zn (1+\zeta) Q},\T{R}{H}}_H \, , \qquad
p_{h_2} = \parz{\frac{\zn (1-\zeta) Q}{2 \sqrt{2}}, \frac{\parz{m^2 +\Tscsq{R}{H}}\sqrt{2}}{\zn (1-\zeta) Q},-\T{R}{H}}_H \, . \label{e.varahadron}
\end{align}
The hadron frame transverse components of $R$ are related to the total invariant mass of the dihadron pair via
\begin{equation}
\Tscsq{R}{H} = \frac{M_h^2 (1 - \zeta^2)}{4} - m^2 \, ,
\label{e.RT2Mh}
\end{equation}
which establishes the range of kinematically allowed $\zeta$, 
\begin{equation}
-\zeta_\text{max} \leq \zeta \leq \zeta_\text{max} \, , \qquad \text{with} \qquad \zeta_\text{max} \equiv \sqrt{1 - \frac{4 m^2}{M_h^2}} \, .
\end{equation}
Therefore, 
to satisfy leading power factorization, one must focus on the limit $4 m^2 \leq \, M_h^2 \ll z Q^2$. 
Note also that if $M_h^2$ is allowed to become too large relative to $m^2$, then the $\zeta \approx \pm1$ region starts to contribute. Then the hadrons in the pair can have very asymmetric rapidities, one of the two hadrons inside the pair can carry a very small fraction of quark momentum, and there could be a danger of contamination by soft hadrons outside of the region where factorization is valid. 
Thus, it is important to restrict consideration to an appropriate range of $M_h^2$, {\textit{e.g.}} by keeping a fixed upper bound on the ratio $M_h^2/m^2$ as $Q \to \infty$.
Another useful relation is 
\begin{equation}
E_{\hv{1}} E_{\hv{2}} = \frac{z_1 z_2 Q^2}{4} + \text{p.s.} = \frac{z^2 Q^2 (1 - \zeta^2)}{16} + \text{p.s.} \, .
\end{equation}
The lab frame components of $\hv{1}$ and $\hv{2}$, expressed in terms of the variables above, are listed in \aref{momentum}. Using \eref{variablesa} and \eref{varahadron}, variable transformations give
\begin{align}
2 E_{\hv{1}} (2 \pi)^3 2 E_{\hv{2}} (2 \pi)^3 \frac{\diff{\cdots}{}}{\diff[3]{\vect{p}_{h_1}}{} \diff[3]{\vect{p}_{h_2}}{}} 
&{}= \frac{32\pi^3}{M_h} 2 E_{h} (2 \pi)^3 \frac{\diff{\cdots}{}}{\diff[3]{\vect{p}_h}{} \diff{M_h}{} \diff{\zeta}{} \diff{\phi_{R,H}}{}} + \text{p.s.} \, , \label{e.varchange1}
 \end{align}
with the left-hand variables in the lab frame, $\phi_{R,H}$ the azimuthal angle of $\T{R}{H}$ in the hadron frame, and with power suppressed errors being dropped at various points on the right-hand side. 
So, \eref{fact_formuladi} becomes
\begin{align}
\label{e.fact_formuladi2}
& \frac{32\pi^3}{M_h} 2 E_{h} (2 \pi)^3 \frac{\diff{\sigma}{}}{\diff[3]{\vect{p}_h}{} \diff{M_h}{} \diff{\zeta}{} \diff{\phi_{R,H}}{}}
= \int_z^1 \frac{\diff{\zetasym}{}}{\zetasym^2} \parz{2 E_{\hat{k}} (2 \pi)^3 \frac{\diff{\hat{\sigma}}{}}{\diff[3]{\vect{\hat{k}}}{}} } d(\zetasym,\zeta,M_h,\phi_{R,H}) + \text{p.s.}  \, , 
\end{align}
where now we have written out the arguments of the dihadron fragmentation function explicitly by defining
\begin{equation}
d(\zetasym,\zeta,M_h,\phi_{R,H}) \stackrel{n=2}{\equiv} d(\zetasym,\left\{ p_h \right\})  \, .
\end{equation}
The new function arguments $\zeta,M_h,\phi_{R,H}$ describe the relative internal momentum of the pair of hadrons. Comparing \eref{fact_formuladi2} with \eref{Yprime} shows that, with this particular choice of variables, the internal phase space of the dihadron pair is
\begin{equation}
\parz{ \frac{\diff{}{}}{\diff{Y'}}}_H \cdots \;\; = \frac{32\pi^3}{M_h} \frac{\diff{}{}}{\diff{M_h}{} \diff{\zeta}{} \diff{\phi_{R,H}}{}} \cdots \, .
\end{equation}
The dihadron fragmentation function in \eref{fact_formuladi2} is still the one from \eref{ffdef}. From the discussion in \sref{ffs},
\begin{equation}
\frac{1}{32\pi^3} \int_V \diff{\zeta}{} M_h  \diff{M_h}{} \diff{\phi_{R,H}}{} d(\zetasym,\zeta,M_h,\phi_{R,H}) 
\label{e.inter}
\end{equation}
is the density (in $\diff{\zetasym}{}$) of dihadron pairs whose relative internal momenta lie within the integration volume $V$. Often one is interested in the cross section differential in $M_h$ and $z$. Then, one may apply all the integrals in \eref{inter} to \eref{fact_formuladi2} apart from $\int \diff{M_h}{}$ to define a reduced dihadron fragmentation function differential in $\xi$ and $M_h$,
\begin{equation}
 d_\text{red,1}(\zetasym,M_h) \equiv 
  \int_{-\zeta_\text{max}}^{\zeta_\text{max}} \diff{\zeta}{} \int_0^{2 \pi}  \diff{\phi_{R,H}}{}  \frac{M_h}{32 \pi^3} d(\zetasym,\zeta,M_h,\phi_{R,H}) \, .
\label{e.extended}
\end{equation}
Note the limits of integration on the $\zeta$ integral, which both approach zero as $M_h \to 2 m$. 
This reduced $d_\text{red,1}(\zetasym,M_h)$ is the density of hadrons with respect to the usual $\diff{\zetasym}{}$ phase space, but now it is also a density with respect to $\diff{M_h}$.\footnote{If we wished, we could have defined still another fragmentation function by absorbing the $M_h/(32 \pi^3)$ of \eref{inter} into $d(\zetasym,\zeta,M_h,\phi_{R,H})$ to make it like a density in its arguments $\zeta,M_h,\phi_{R,H}$. We refrain from doing this here to make keeping track of Jacobian factors simpler.}
Applying the $\int M_h \frac{\diff{\zeta}{} \diff{\phi_{R,H}}{}}{32 \pi^3}$ integral to the cross section in \eref{fact_formuladi2} gives
\begin{align}
\label{e.fact_formuladi3}
2 E_h (2 \pi)^3 \frac{\diff{\sigma}{}}{ \diff[3]{\vect{p}_h}  \diff{M_h}{}}  
 = \int_z^1 \frac{\diff{\zetasym}{}}{\zetasym^2} \parz{2 E_{\hat{k}} (2 \pi)^3 \frac{\diff{\hat{\sigma}}{}}{\diff[3]{\vect{\hat{k}}}{}} } d_\text{red,1}(\zetasym,M_h) 
 + \text{p.s.} \, .
\end{align}
This form of the factorization formula matches exactly the single hadron case apart from the cross section now being differential in $M_h$ and the fragmentation function being a density in both total momentum fraction $\zetasym$ and the dihadron mass $M_h$. All subsequent steps to simplify the cross section are identical to those for a single-hadron fragmentation function, apart from the need to carry along an $M_h$ argument in the fragmentation function. The steps in \eref{diffzsimp} are also unchanged and give
\begin{align}
\frac{\diff{\sigma}{}}{\diff{z}{} \diff{M_h}{}} &= \frac{4 \pi \alpha^2}{3 Q^2} d_\text{red,1}(z,M_h) 
+ \text{p.s.}
+ \order{\alpha_s} = \sigma_0 d_\text{red,1}(z,M_h) 
+ \text{p.s.} 
+ \order{\alpha_s} \, .\label{e.dzcrosssectionndihadron}
\end{align}
With the error terms dropped, this is the form of the cross section that is commonly used for phenomenological applications (see Eq.~(9) of Ref.~\cite{Courtoy:2012ry}).\footnote{Recall that here we use a single parton with unit charge, hence 
we must restore the overall factor of $N_c = 3$ and the squared fractional charge $e_q^2$ with a sum over flavors to get back to the full QCD version in Ref.~\cite{Courtoy:2012ry}.} 
There, it is based on the operator definition in Eq.~(11) from Ref.~\cite{Bianconi:1999cd} (or Eq.~(16) from Ref.~\cite{Bacchetta:2003vn}), which matches \eref{tmdgraphical} above (see \aref{JAM-Pavia2} for a more detailed comparison). 
Thus, the definitions used in these earlier applications are consistent with the factorization derivation.

To reduce the cross section to a still simpler form, one may integrate over $M_h$. Then, 
\begin{align}
\frac{\diff{\sigma}{}}{\diff{z}{}} &= \frac{4 \pi \alpha^2}{3 Q^2} d_\text{red,2}(z;M_{h,\text{max}}) 
+ \text{p.s.} 
+ \order{\alpha_s} 
= \sigma_0 d_\text{red,2}(z;M_{h,\text{max}}) 
+ \text{p.s.} 
+ \order{\alpha_s}\, , \label{e.dzcrosssectionndihadron2}
\end{align}
where 
\begin{equation}
d_\text{red,2}(z;M_{h,\text{max}}) \equiv \int_{2 m}^{M_{h,\text{max}}} \diff{M_h}{} \, d_\text{red,1}(z,M_h) \, .
\end{equation}
The $M_h$ integral is ultraviolet divergent in general, so the cutoff $M_{h,\text{max}}$ cannot be made arbitrarily large. To maintain factorization one must ensure that $M_{h,\text{max}}^2/z \ll Q^2$. Thus, the fully integrated dihadron fragmentation function comes with an additional auxiliary ultraviolet scale $M_{h,\text{max}}$. 

In a simple parton model treatment, one may expect that integrating over $z$ down to a small $z$ cutoff and using the definition of the inclusive cross section gives an approximate relation analogous to \eref{partonmodel}, 
\begin{align}
\underset{\text{types}}{\sum_{\text{hadron}}} 
\int_{z_\text{min}}^1 \diff{z}  d_\text{red,2}(z;M_{h,\text{max}};\mu) \approx \langle N_2 \rangle (z_\text{min},M_{h,\text{max}},\mu) \, ,\label{e.partonmodel2}
\end{align}
where $\langle N_2 \rangle (z_\text{min},M_{h,\text{max}},\mu)$ is a multiplicity of dihadron pairs inside a region specified by $M_{h,\text{max}}$ and $z_\text{min}$, and we have restored the explicit scale parameter $\mu$ as an argument. The auxiliary scales $\mu$, $M_{h,\text{max}}$ and $z_\text{min}$ may be thought of as defining the borders of a jet or jet-like structure. The $z_\text{min}$ should be large relative to $\Lambda^2/Q^2$ and $M_{h,\text{max}}$ should be much less than $Q$. The ``$\approx$'' refers to both neglected subleading powers and higher orders in $\alpha_s$. If the total number of hadrons in this region is $N$, then the number of dihadron pairs is $N \choose 2$ multiplied by the $2!$ ways of relabeling the hadrons. Thus, $\langle N_2 \rangle = \langle N(N-1) \rangle$.

The general \emph{zeroth order} dihadron factorization formula is 
\begin{align}
\label{e.fact_formula_zeroth}
2 E_{h_1} (2 \pi)^3 2 E_{h_2} (2 \pi)^3 \frac{\diff{\sigma}{}}{\diff[3]{\vect{p}_{h_1}}{} \diff[3]{\vect{p}_{h_2}}{}} &= \int_z^1 \frac{\diff{\zetasym}{}}{\zetasym^2} \parz{2 E_{\hat{k}} (2 \pi)^3 \frac{\diff{\hat{\sigma}}{}}{\diff[3]{\vect{\hat{k}}}{}} } d(\zetasym,\left\{p_h\right\}) + \text{p.s.} \, \no
 &= \frac{8 (2 \pi)^3}{Q^2} \int_z^1 \frac{\diff{\zetasym}{}}{\zetasym^2} \parz{ \frac{\diff{\hat{\sigma}}{}}{\hat{z} \diff{\hat{z}} \diff{\Omega}} } d(\zetasym,\left\{p_h\right\}) + \text{p.s.} \no
 &= \frac{8 (2 \pi)^3}{Q^2} \int_z^1 \frac{\diff{\zetasym}{}}{\zetasym^2} \parz{ \frac{\alpha^2}{2 Q^2} \left[ \widehat{F}_1 + \widehat{F}_2 \frac{\hat{z}}{4} \sin^2 \theta \right] } d(\zetasym,\left\{p_h\right\}) + \text{p.s.} \no
  &= \frac{8 (2 \pi)^3}{Q^2}  \frac{1}{z} \parz{ \frac{\alpha^2}{2 Q^2} \left[ 1 - 2 \frac{1}{4} \sin^2 \theta \right] } d(z,\left\{p_h\right\}) + \text{p.s.} + \order{\alpha_s} \no
  &= \frac{2 \alpha^2 (2 \pi)^3}{z Q^4} \parz{  1 +  \cos^2 \theta } d(z,\left\{p_h\right\}) + \text{p.s.} + \order{\alpha_s} \, ,
\end{align}
where we used the expressions for $\widehat{F}_{1,2}$ in Eq.~\eqref{e.hatF1F2}. When applying \eref{fact_formula_zeroth} phenomenologically, the exact choice of variables one uses for $\vect{p}_{h_1}$ and $\vect{p}_{h_2}$ is an arbitrary convention unrelated to the basic definition of the dihadron fragmentation function $d(z,\left\{p_h\right\})$ or its interpretation. The choice we used above that leads to \eref{fact_formuladi2} is only one possible example. 

\subsection{Variable choice B}

Another choice of variables, used in Ref~\cite{Majumder:2004wh}, is the set of three-vector momenta
\begin{equation}
\vect{p}_h = \vect{p}_{h_1} + \vect{p}_{h_2} \, , \qquad \vect{l} = \vect{p}_{h_1} - \vect{p}_{h_2} \, ,
\end{equation}
with the relevant $\vect{l}$ variables being its transverse components and azimuthal angle in the hadron frame, and with the cross section made differential in $z_1$ and $z_2$ separately.
In Ref.~\cite{Majumder:2004wh}, $\vect{l}$ was called $\vect{q}$, but we use a different label here to avoid overlapping with the virtual photon momentum label. 
The components of the hadron momenta in the laboratory frame are listed in \aref{momentum}.
Below, $\T{l}{}$ is the transverse part of $\vect{l}$ in the hadron frame and $\phi_T$ is its hadron frame azimuthal angle. Since our focus is the SIA factorization region, we consider only the case where the angle between $\vect{p}_{h_{1,2}}$ and $\vect{p}_h$ is less than 90$^o$, i.e. we keep only positive square roots. The invariant mass squared $M_h^2$ is then
\begin{align}
M_h^2 \to M_h^2(Q,z_1,z_2,\Tsc{l}{}) =  \frac{z_1 z_2 Q^2}{2} \parz{1 + \frac{\Tscsq{l}{}+ 4 m^2}{z_1 z_2 Q^2} - \sqrt{1 - \frac{\Tscsq{l}{} + 4 m^2}{z_1^2 Q^2}} \sqrt{1 - \frac{\Tscsq{l}{} + 4 m^2}{ z_2^2 Q^2}}  } = \frac{\Tscsq{l}{} z^2}{4 z_1 z_2} + \text{p.s.} \, , 
\end{align}
where $\Tsc{l}{}$ is the magnitude of the transverse part of $\vect{l}$ in the hadron frame. We may change variables as follows, 
\begin{align}
\diff[3]{\vect{p}_{h_1}}{} \diff[3]{\vect{p}_{h_2}}{} &= 
\diff{\phi_T} \diff{\phi} \diff{\theta} \diff{\Tsc{l}{}} \diff{|\vect{p}_{h_1}|} \diff{|\vect{p}_{h_2}|} \frac{\Tsc{l}{} \sin \theta}{4} \parz{|\vect{p}_{h_1}| + |\vect{p}_{h_2}|}^2 + \text{p.s.} \, , \no
&{}= \diff{\phi_T} \diff{\phi} \diff{\cos(\theta)} \diff{\Tscsq{l}{}} \diff{|\vect{p}_{h_1}|} \diff{|\vect{p}_{h_2}|} \frac{1}{8} \parz{|\vect{p}_{h_1}| + |\vect{p}_{h_2}|}^2 + \text{p.s.} \, . \label{e.firstjac}
\end{align}
Now the ``$\text{p.s.}$'' includes $\Tsc{l}{}/Q$-suppressed contributions. 
Equation~\eqref{e.firstjac} agrees with the Jacobian factor listed on page 10 of 
Ref.~\cite{Majumder:2004wh}. 
Using $|\vect{p}_{h_1}| = E_{h_1} = z_1 Q/2$ + \text{p.s.}, $|\vect{p}_{h_2}| = E_{h_2} = z_2 Q/2 + \text{p.s.}$, this is also  
\begin{align}
\diff{\phi_T} \diff{\phi} \diff{\cos(\theta)} \diff{\Tscsq{l}{}} \diff{|\vect{p}_{h_1}|} \diff{|\vect{p}_{h_2}|} \frac{1}{8} \parz{|\vect{p}_{h_1}| + |\vect{p}_{h_2}|}^2 &{}= \frac{1}{32} \diff{\phi_T} \diff{\Omega} \diff{\Tscsq{l}{}} \diff{|\vect{p}_{h_1}|} \diff{|\vect{p}_{h_2}|} Q^2 \parz{z_1 + z_2}^2 
 \, \no
&{}= \frac{z^2}{128} \diff{\phi_T} \diff{\Omega} \diff{\Tscsq{l}{}} \diff{z_1} \diff{z_2} Q^4  \, , 
\end{align}
all up to power-suppressed errors. Then
\begin{align}
\diff{Y}{} = \frac{\diff[3]{\vect{p}_{h_1}}{} \diff[3]{\vect{p}_{h_2}}{}}{2 E_{h_1} (2 \pi)^3 2 E_{h_2} (2 \pi)^3}  &{}= \frac{Q^2 z^2}{128 z_1 z_2 (2 \pi)^6} \diff{\Omega} \diff{\phi_T} \diff{\Tscsq{l}{}} \diff{z_1} \diff{z_2} + \text{p.s.} \, .
\end{align}
Now \eref{withstructfuncts} becomes 
\begin{align}
\frac{\diff{\sigma}}{\diff{Y}} &= {2 E_{h_1} (2 \pi)^3 2 E_{h_2} (2 \pi)^3} \frac{\diff{\sigma}}{\diff[3]{\vect{p}_{h_1}}{} \diff[3]{\vect{p}_{h_2}}{}} \no
&= \frac{128 z_1 z_2 (2 \pi)^6}{Q^2 z^2} \frac{\diff{\sigma}}{\diff{\Omega} \diff{\phi_T} \diff{\Tscsq{l}{}} \diff{z_1} \diff{z_2}} = \frac{4 (2 \pi)^3 \alpha^2}{Q^4} \left[ F_1(z,Q^2,\left\{p_h\right\}) + \frac{z}{4} F_2(z,Q^2,\left\{p_h\right\}) \sin^2 \theta \right] \, ,\label{e.dz1dz2crosssection}
\end{align}
or 
\begin{align}
  \frac{\diff{\sigma}}{\diff{\Omega} \diff{\phi_T} \diff{\Tscsq{l}{}} \diff{z_1} \diff{z_2}} = \frac{\alpha^2 z^2}{32 (2 \pi)^3 z_1 z_2 Q^2} \left[F_1(z,Q^2,\left\{p_h\right\}) + \frac{z}{4} F_2(z,Q^2,\left\{p_h\right\}) \sin^2 \theta \right] \, . \label{e.dz1dz2crosssection2}
\end{align}
Integrating over $\diff{\Omega}$,
\begin{align}
  \frac{\diff{\sigma}}{\diff{\phi_T} \diff{\Tscsq{l}{}} \diff{z_1} \diff{z_2}} = \frac{\alpha^2 z^2 \pi}{16 (2 \pi)^3 z_1 z_2 Q^2} \left[ 2 F_1(z,Q^2,\left\{p_h\right\}) + \frac{z}{3} F_2(z,Q^2,\left\{p_h\right\}) \right]  \, . \label{e.dz1dz2crosssection3}
\end{align}
The general factorized expressions for $F_1(z,Q^2)$ and $F_2(z,Q^2)$ are still just \erefs{fact_formulaF1}{fact_formulaF2}, with exactly the same definition \eref{ffdef} (with \eref{collinearoperator}) for the fragmentation function as in the single hadron case. Now the ``\text{p.s.}'' contributions include terms that vanish like $\Tscsq{l}{}/Q^2$. 

If one intends to integrate over some range of $\Tscsq{l}{}$ and $\phi_T$ in an experiment, then it may be convenient to define a third type of reduced fragmentation function that absorbs the kinematical Jacobian factors from the cross section above, 
\begin{equation}
 d_\text{red,3}(\zetasym x_1, \zetasym x_2,z) \equiv 
  \frac{z}{16 (2 \pi)^3 z_1 z_2} \int_V \diff{\phi_T}{} \diff{\Tscsq{l}{}}{} \, d(\zetasym x_1, \zetasym x_2, l_T, \phi_T) \, ,
\label{e.extended3}
\end{equation}
where, recalling the definitions of $x_{1,2}$ in \eref{xdefs}, we have now defined 
\begin{equation}
d(\zetasym x_1, \zetasym x_2, l_T, \phi_T) \stackrel{n=2}{\equiv} d(\zetasym,\left\{ p_h \right\}) \label{e.argumentdef}
\end{equation}
in these variables. The reason that we have defined the function arguments as in \eref{argumentdef} is that universality means the fragmentation function only depends on the parton momentum and the observed hadron momentum. Then, boost invariance means the longitudinal components enter only in ratios of the plus momentum components. These consist of all combinations of $x_1$, $x_2$, and $\zetasym$, but $x_1 + x_2 = 1$ so all three momentum fractions are determined from knowledge of $\zetasym x_1$ and $\zetasym x_2$. When $\xi \to z$, these just become $z_1$ and $z_2$.
The integration region needs to be restricted to $\Tscsq{l}{} \ll Q^2$ and fixed $z_1$ and $z_2$ away from 0 to preserve factorization. Integrating over $\Tsc{l}{}$ and $\phi_T$, \eref{dz1dz2crosssection3} becomes,
\begin{equation}
\frac{\diff{\sigma}{}}{\diff{z_1}{} \diff{z_2}{} } = \frac{\pi \alpha^2 z}{Q^2} \left[ 2 F_1(z_1,z_2,Q^2) + \frac{z}{3} F_2(z_1,z_2,Q^2) \right] \, + \text{p.s.} \, \label{e.dz1dz2crosssection5}
\end{equation}
with these structure functions obeying the same factorization formulas 
\begin{align}
\label{e.fact_formulaF1dz1dz2}
F_1(z_1,z_2,Q^2) &{}= \int \frac{\diff{\zetasym}{}}{\zetasym^2} \widehat{F}_1(\hat{z},Q^2) \, d_\text{red,3}(\zetasym x_1, \zetasym x_2, z)+ \text{p.s.} \, , \\
\label{e.fact_formulaF2dz1dz2}
F_2(z_1,z_2,Q^2) &{}= \int \frac{\diff{\zetasym}{}}{\zetasym^3} \widehat{F}_2(\hat{z},Q^2) \, d_\text{red,3}(\zetasym x_1, \zetasym x_2, z)+ \text{p.s.} \, .
\end{align}
That is, they are again the same as in the single hadron formula in \eref{dzcrosssection}, but now with the fragmentation functions integrated over the (restricted range of) physical phase space in \eref{extended3}. Recall that to keep the power suppressed errors from growing too large, one needs to maintain a lower bound on the possible values of $z_1$ and $z_2$.

Using the zeroth order parton model level expressions for $F_1$ and $F_2$ in \eref{dz1dz2crosssection3} gives
\begin{align}
   \frac{\diff{\sigma}}{\diff{\phi_T} \diff{\Tscsq{l}{}} \diff{z_1} \diff{z_2}} &= \frac{\alpha^2 z \pi}{4 (2 \pi)^3 z_1 z_2 Q^2 3} d(z,\left\{ p_h \right\}) + \order{\alpha_s} + \text{p.s.} \no
  &=  
  \frac{4 \pi \alpha^2}{3 Q^2} \frac{z}{16 (2 \pi)^3 z_1 z_2} d(z,\left\{ p_h \right\}) + \order{\alpha_s} + \text{p.s.}  \no
  &= \sigma_0 \frac{z}{16 (2 \pi)^3 z_1 z_2} d(z,\left\{ p_h \right\}) + \order{\alpha_s} + \text{p.s.}  \, . \label{e.dz1dz2crosssection4}
\end{align}

To recover the specific forms of the expressions in Ref.~\cite{Majumder:2004wh} we must note the following:
\begin{itemize}
\item In Ref.~\cite{Majumder:2004wh}, the identification of physical momentum components with internal partonic momentum fractions is imposed from the very beginning. Therefore, to match expressions we must take $\zetasym = z$, $\zetasym_1 = z_1$, etc., from the outset. 
\item From Eqs.~(24,27) of Ref.~\cite{Majumder:2004wh}, the trace in the integrand of our  \eref{collinearoperator} above is $2 p_{h,H}^+ T$, where $T$ is the quantity defined in Ref.~\cite{Majumder:2004wh}.
\end{itemize}
It is also useful to note that 
\begin{equation}
2 p_{h, H}^+ \int \diff{k_H^+} \delta \parz{k_H^+ - p_{h,H}^+/\zetasym} \ldots = 2 \zetasym^2 \int \diff{k_H^+} \delta \parz{\zetasym- \frac{p_{h,H}^+}{k_H^+}} \ldots \, .
\end{equation}
Then \eref{extended3} matches Eq.~(37) of Ref.~\cite{Majumder:2004wh} when we observe that, 
\begin{align}
d_\text{red,3}(\zetasym x_1, \zetasym x_2,z) &{}= 
\frac{z}{16 (2 \pi)^3 z_1 z_2} \int_V \diff{\phi_T}{} \diff{\Tscsq{l}{}}{} 2 p_{h,H}^+ \frac{\zetasym}{4} \int^{\sim \mu^2} \frac{\diff{k^-_H} \diff[2]{\T{k}{H}}}{(2 \pi)^4} {\rm Tr} \left[\gamma^+  T \right]  {\Big \vert}_{k_H^+ = p_{h,H}^+/\xi} \no
&{}= \frac{z}{32 (2 \pi)^3 z_1 z_2} \int_V \diff{\phi_T}{} \diff{\Tscsq{l}{}}{} \zetasym^3 \int^{\sim \mu^2} \frac{\diff{k^+_H}{} \diff{k^-_H} \diff[2]{\T{k}{H}}}{(2 \pi)^4} {\rm Tr} \left[\gamma^+  T \right] \delta\parz{\xi - p_{h,H}^+/k_H^+} \no
&{}\to \frac{z^4}{32(2\pi)^2 z_1 z_2} \int_V \diff{\Tscsq{l}{}}{} \int \frac{\diff[4]{k}}{(2\pi)^4} {\rm Tr} \left[\gamma^+  T \right] \delta\parz{\xi - p_{h,H}^+/k_H^+} = d_\text{red,3}(z_1,z_2,z) \, . \label{e.MWform}
\end{align}
On the last line, we have made the replacement $\xi \to \zn \approx z$ in all parts of this reduced fragmentation function apart from the $\delta$-function, and we have integrated over $\phi_T$.\footnote{It is important to keep in mind, however, that the identification of the parton momentum fraction $\xi$ with $z$ fails beyond zeroth order in the hard part.} 

The same integration that is used in the reduced fragmentation function of \eref{extended3} when applied to the zeroth order parton model level cross section in \eref{dz1dz2crosssection4} gives 
\begin{equation}
\frac{\diff{\sigma}{}}{\diff{z_1} \diff{z_2}} = \sigma_0 d_\text{red,3}(z_1,z_2,z) + \text{p.s.} + \order{\alpha_s} \, .
\end{equation}
This recovers Eq.(36) of Ref.~\cite{Majumder:2004wh}, which was obtained using the reduced fragmentation function in Eq.(37) of Ref.~\cite{Majumder:2004wh} (and \eref{extended3} above). 

An important point to note here is that the kinematical $1/(z_1 z_2)$ factor in \eref{MWform} is unrelated to the $1/(4 \zetasym)$ in the basic operator definition \eref{ffdef} of the fragmentation function. The latter arises from the parton momentum while the former is associated with the hadron momenta and is a consequence of the specific and arbitrary set of variables with which we express the SIA cross section. That combination of kinematical factors does not arise if, for example, the variables are $z$ and $M_h$ as in \sref{varA}.

Reference~\cite{Majumder:2004wh} also includes a treatment of evolution, although it is not obtained from the renormalization of a specific operator definition but rather from the reabsorption of collinear divergent logarithms from higher order perturbation theory diagrams. It also takes into account contributions from 2-parton-to-2-hadron fragmentation mechanisms, which are beyond the scope of this paper.  
Therefore, comparing the evolution treatment there with the operator definition oriented discussion here is not straightforward, and we leave such comparisons to future work.

\section{Alternative definitions with nonuniversal prefactors} 
\label{s.alternative}

It was recently argued in Ref.~\cite{Pitonyak:2023gjx} 
(see also Refs.~\cite{metztalk,pittalk} for additional details) 
that the basic operator definition for the fragmentation function in \eref{ffdef} needs to be modified, when the final state is a dihadron, to 
\begin{align}
& d\parz{\zetasym,-\zetasym \T{k}{H},\hv{1},\hv{2}} \, \no
& \qquad \equiv  \frac{1}{64 \pi^3 \zetasym_1 \zetasym_2}  {\rm Tr} \SumInt_X \int \frac{\diff{x^-}{} \diff[2]{\T{x}{}}{}}{(2 \pi)^3} e^{i x^- p_{h,H}^+/\zetasym - i \T{k}{H} \cdot \T{x}{} } \langle 0 | \gamma^+ \psi(x/2) |\hv{1},\hv{2}, X \rangle \langle \hv{1},\hv{2}, X |\overline{\psi}(-x/2)| 0 \rangle \, \no
& \qquad \equiv \frac{1}{64 \pi^3 \zetasym^2 x_1 x_2} {\rm Tr} \SumInt_X \int \frac{\diff{x^-}{} \diff[2]{\T{x}{}}{}}{(2 \pi)^3} e^{i x^- p_{h,H}^+/\zetasym - i \T{k}{H} \cdot \T{x}{}}  \langle 0 | \gamma^+ \psi(x/2) |\hv{1},\hv{2}, X \rangle \langle \hv{1},\hv{2}, X |\overline{\psi}(-x/2)| 0 \rangle \, . 
\label{e.ffdefmod}
\end{align}
As before, $x = (0,x^-,\T{x}{})$. The $x_1 x_2$ factor only depends on the external kinematics of the hadrons, so apart from such constant factors there is an extra overall $1/\zetasym$ prefactor in this definition relative to standard definition.\footnote{Here we are interpreting the prefactors as referring to momentum fractions $\xi_1$ and $\xi_2$, as is stated in Ref.~\cite{Pitonyak:2023gjx}, rather than kinematical variables $z_1$ and $z_2$. If the latter was the case, then the definition would still differ from the standard one, but by a factor of $\xi$ rather than $1/\xi$. It would also acquire additional process dependence since $z_1$ and $z_2$ are defined in terms of the external momentum $q$.} The corresponding collinear fragmentation function is deduced by further applying the transverse parton momentum integral
to it. For each $n > 2$ case, an analogous definition is proposed in Ref.~\cite{Pitonyak:2023gjx}, with an extra prefactor of $1/\zetasym_i$ multiplying the basic definition for each additional hadron $i$ that appears in the $n$-hadron. Other Jacobian factors are then used later in Ref.~\cite{Pitonyak:2023gjx} to convert the definition in \eref{ffdefmod} into functions that are to be integrated over specific momentum variables. 

Equation~\eqref{e.ffdefmod} contradicts the standard fragmentation function definition that follows from steps like those in \srefs{fact}{ffs}, and since this impacts the interpretation of past phenomenological work it is important to understand the origin of the discrepancy (see Ref.~\cite{Rogers:2024zvi} for additional discussions of this). 

We identify a combination of two main reasons that Ref.~\cite{Pitonyak:2023gjx} arrives at a different definition: 
\begin{enumerate}
\item Reference~\cite{Pitonyak:2023gjx} conflates parton momentum and hadron momentum, and in particular the parton momentum fractions that we call $\zetasym_1$, $\zetasym_2$, and $\zetasym$ are identified with the external kinematical variables like $z_1$, $z_2$, and $z$. While these are fixed by a $\delta$-function to be numerically equal in zeroth order hard scattering, they have different roles in factorization and in operator definitions for fragmentation functions. In higher orders, and also in the treatment of evolution, momentum fractions are not generally equal to external kinematical variables. 
\item The treatment in Ref.~\cite{Pitonyak:2023gjx} infers the form of the dihadron fragmentation function from the requirement that an incorrect sum rule of the type
\begin{equation}
\underset{\text{types}}{\sum_{\text{hadron}}} 
\int \diff{\zetasym_1} \diff{\zetasym_2} \diff[2]{\T{p}{p{h_1}}} \diff[2]{\T{p}{p{h_2}}}  \ldots d\parz{\zetasym,\T{p}{p{h}},\left\{ \hv{n} \right\}} = \langle N_n \rangle  \label{e.sumrule}
\end{equation}
must be preserved, where $\langle N_n \rangle$ is a multiplicity of observed final state $n$-hadrons. The specific argument of Ref.~\cite{Pitonyak:2023gjx} is that the normalization factors of $ 2 E_i (2 \pi)^3$ for each observed hadron $i$ in the
asymptotic $n$-hadron state,  
combined with \eref{sumrule}, would imply that the operator definition of multihadron fragmentation functions must come with extra additional prefactors of $1/(2 E_i (2 \pi)^3)$.  
After changing variables in \eref{sumrule}, and with the conflation of momentum fractions and kinematical variables remarked upon in the first item above, that would seem to require modified prefactors like the one in \eref{ffdefmod}.

However, those external kinematical prefactors are already accounted for on the left-hand side of the basic expression for the cross section in \eref{fact_formula} and in \eref{fact_formula_zeroth}. Repeating them in the definitions of fragmentation functions themselves amounts to overcounting them. The $1/(4 \zetasym)$ prefactor in the standard definition, \eref{ffdef}, actually arises from the partonic momentum, not from the phase space of observed hadrons.  

Equation~\eqref{e.sumrule} itself fails for several reasons: In QCD, it fails for the interesting and rather fundamental reason that hadronic final states have no overlap with states that have partonic quantum numbers. For a recent discussion of this and related issues see Ref.~\cite{Collins:2023cuo}. More generally, the sum rule fails even in nongauge theories because it requires integrating the momenta outside the region where factorization is valid, and where the right-hand side of \eref{sumrule} is not meaningful. Indeed, the integral in~\eqref{e.sumrule} is generally ultraviolet divergent in theories that require renormalization.  Recalling the discussions from \sref{interps} above, however, it is important to emphasize that preserving such a sum rule is not necessary for the standard fragmentation function definition to retain a parton model interpretation, so long as the usual caveats are accounted for.

\end{enumerate}

If one does use a definition like \eref{ffdefmod} in a factorization formula then, as the steps shown in previous sections illustrate, one is forced to modify the hard partonic part with extra powers of $\zetasym$ for 
each additional hadron that appears in the observed $n$-hadron state.  
This removes the property expected for a satisfactory factorization theorem that hard parts are independent of the details of external  nonperturbative states.  

Moreover, if one modifies the basic definition of fragmentation functions with nonuniversal $\zetasym$-dependent prefactors this 
also modifies the renormalization factors in the renormalized fragmentation functions, which unavoidably changes the evolution kernels. To see this, recall how a renormalized fragmentation function is related to its bare operator definition (see, e.g., Sec. 12.2.2 in Ref.~\cite{Collins:2011qcdbook}),
\begin{equation}
\label{e.renorm}
d_{h/j}(\zetasym,\left\{p_h\right\};\mu) = \lim_{\epsilon \to 0} \sum_{j'} \int_z^1 \frac{\diff{\zetasym'}}{\zetasym'} d_{(0)h/j'} (\zetasym/\zetasym',\{p_h\}) Z_{j'j}(\zetasym',g(\mu),\epsilon) = \sum_{j'} d_{(0)h/j'} \otimes Z_{j'j} \, .
\end{equation}
Here we have restored parton flavor subscripts and the explicit $\mu$ and $\epsilon$ arguments that signal ultraviolet regularization and renormalization in the $\msbar$ scheme. Also, the ``$(0)$'' subscripts on the right-hand side denote bare operators. The $d_{(0)h/j'} (\zetasym/\zetasym',\{p_h\})$ is given by the definition in \sref{ffs}, but now we specify bare operators for the fields. 
The renormalization factor $Z_{j'j}(\zetasym',g(\mu),\epsilon)$ is independent of the identity of the hadronic state $h$, and this gives rise to the hadron-independence of evolution kernels. Indeed, a standard method for calculating $Z_{j'j}$ in perturbation theory is to exploit the $h$-independence by replacing $h$ with an on-shell partonic state. This then allows for simple calculations of the evolution kernels directly from massless, on-shell Feynman diagrams. That procedure is invalidated, however, if the overall prefactor in \eref{ffdef} is made dependent (in a $\zetasym$-dependent way) on the identity of $h$, and in particular on the number of hadrons that make up $h$. 

A fast way to see this is with sample calculations in model field theories like a scalar Yukawa theory with nonzero masses and with the interaction terms 
\begin{equation}
- g \overline{\psi} \psi \phi - \frac{\lambda}{3!} \phi^3 \, 
\label{e.yukawaterms}
\end{equation}
in the Lagrangian density.
Here, $\psi$ is a Dirac ``quark'' and $\phi$ is a scalar ``hadron.'' The lowest order hadron-in-quark and dihadron-in-quark fragmentation functions are given by the graphs in \fref{yukawa_unfact}. 
The only contribution to the renormalization factor $Z_{j'j}$ in the collinear fragmentation function at this order of perturbation theory comes from the ultraviolet divergence at large $\Tsc{k}{H}$. In the (di)hadron-in-quark TMD fragmentation functions, the ultraviolet large $\Tsc{k}{H}$ parts refactorize in the following way,  
\begin{equation}
d_{(0),\phi/\phi} \otimes  \mathcal{H}_{\phi/q} \qquad  \text{and} \qquad d_{(0),\phi \phi/\phi} \otimes  \mathcal{H}_{\phi/q}
\end{equation}
where $\mathcal{H}_{\phi/q}$ is the same large-$\Tsc{k}{H}$ hard factor in both cases (see \fref{yukawa_fact}), and $d_{(0),\phi/\phi}$ and $d_{(0),\phi \phi/\phi}$ are zeroth order hadron-in-hadron and $\order{\lambda^2}$ dihadron-in-hadron collinear fragmentation functions, respectively. The steps for refactorizing are exactly analogous to the steps to factorize the cross section itself. The same renormalization factor $Z_{j'j}$ arises for both the single and dihadron cases, but only if both use the same operator definitions.
If the prefactor in the operator definition becomes $h$-dependent in a nontrivial way (e.g., dependent on the number of hadrons in the cluster), then so does $Z_{j'j}$, and therefore so does the evolution kernel.
An extended version of this argument is provided in \aref{yukawa}.
\begin{figure}[h!]
\centering
\includegraphics[width=12cm]
{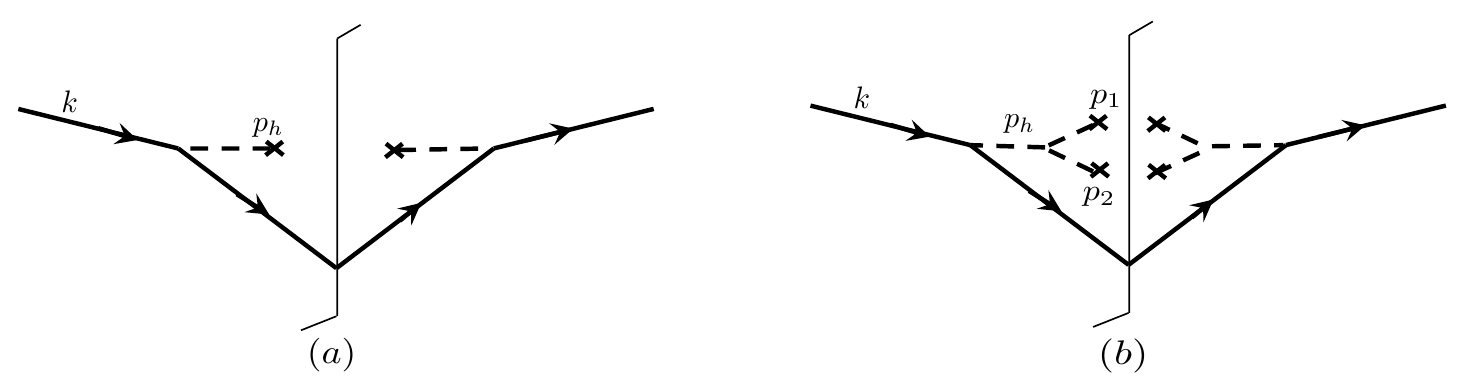}
\caption{Graphs contributing to a single hadron (a) and a dihadron (b) fragmentation function in the Yukawa theory of \eref{yukawaterms}. The integrals and Dirac traces, etc., in \erefs{ffdef}{collinearoperator} are left implicit.}
\label{f.yukawa_unfact}
\end{figure}
\begin{figure}[h!]
\centering
\includegraphics[width=12cm]
{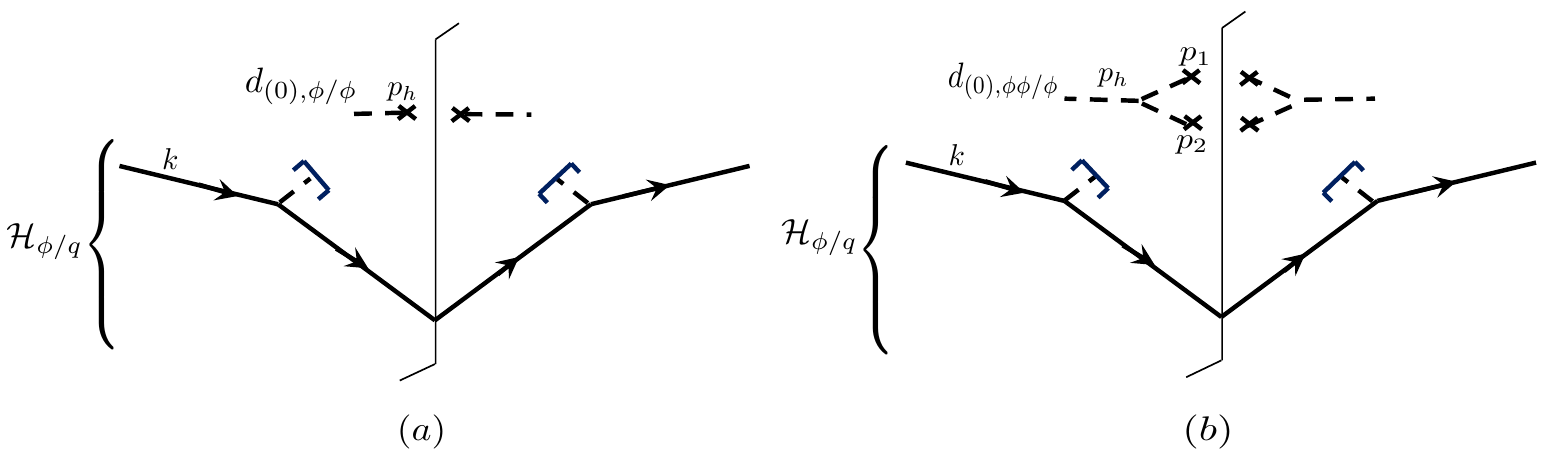}
\caption{Figure \ref{f.yukawa_unfact} with the large $\Tsc{k}{H}$ parts shown factored from collinear fragmentation functions. The purpose of the graphs is to emphasize that the factorized hard contributions from the large $\Tsc{k}{H}$ regions are independent of whether the observed state is a hadron or a small-mass multihadron.}
\label{f.yukawa_fact}
\end{figure}

We acknowledge that Ref.~\cite{Pitonyak:2023gjx} contains an assertion that \eref{dzcrosssectionndihadron} from this paper and Eq.~(9) from Ref.~\cite{Courtoy:2012ry} are derivable, along with the standard evolution equations, starting from the definition given in \eref{ffdefmod}, but we are unable to retrace the steps based on the information provided there. If we use \eref{ffdefmod}, then we instead find that at least one extra factor of $\zetasym$ needs to be absorbed into the hard part to compensate for the extra $1/\zetasym$ in the modified fragmentation function (along with whatever less interesting constant factors like $(2 \pi)^3$ might appear). In that case, \eref{fact_formuladi} becomes
\begin{equation}
\label{e.fact_formuladi2ppcmps}
2 E_{\hv{1}} (2 \pi)^3 2 E_{\hv{2}} (2 \pi)^3 \frac{\diff{\sigma}{}}{\diff[3]{\vect{p}_{h_1}}{} \diff[3]{\vect{p}_{h_2}}{}} = \int_z^1 \frac{\diff{\zetasym}{}}{\zetasym^2} \parz{2 E_{\hat{k}} (2 \pi)^3 \zetasym \frac{\diff{\hat{\sigma}}{}}{\diff[3]{\vect{\hat{k}}}{}} } d_\text{mod}(\zetasym,\left\{ p_h \right\}) 
+ \text{p.s.} \, ,
\end{equation}
where $d_\text{mod}(\zetasym,\left\{ p_h \right\})$ is a dihadron fragmentation function modified by an extra $1/\zetasym$ prefactor. In the limit of the zeroth order parton model approximation this becomes 
\begin{align}
\label{e.fact_formula_zerothppcmps}
2 E_{h_1} (2 \pi)^3 2 E_{h_2} (2 \pi)^3 \frac{\diff{\sigma}{}}{\diff[3]{\vect{p}_{h_1}}{} \diff[3]{\vect{p}_{h_2}}{}} 
  &= \frac{2 \alpha^2 (2 \pi)^3}{Q^4} \parz{  1 +  \cos^2 \theta } d_{\text{mod}}(z,\left\{p_h\right\}) + \text{p.s.} + \order{\alpha_s} \, ,
\end{align}
so there is at least an extra overall factor of $z$ on the right-hand side relative to the normal expression in \eref{fact_formula_zeroth}. To our knowledge, such a formula has not been used in phenomenological applications.

From \eref{fact_formuladi2ppcmps}, retracing the steps from \erefs{varchange1}{dzcrosssectionndihadron} gives the analog of \eref{fact_formuladi3}:
\begin{align}
\label{e.fact_formuladi3b}
2 E_h (2 \pi)^3 \frac{\diff{\sigma}{}}{ \diff[3]{\vect{p}_h}  \diff{M_h}{}}  
 = \int_z^1 \frac{\diff{\zetasym}{}}{\zetasym^2} \parz{2 E_{\hat{k}} \zetasym (2 \pi)^3 \frac{\diff{\hat{\sigma}}{}}{\diff[3]{\vect{\hat{k}}}{}} } d_\text{mod,red,1}(\zetasym,M_h) 
 + \text{p.s.} \, .
\end{align}
Using \erefs{diffzomega1}{partontrans} and following the same steps as in \eref{diffzsimp} then gives
\begin{align}
\frac{\diff{\sigma}{}}{\diff{z}{} \diff{M_h}{}} &=z \int d\Omega \int_z^1 \frac{d\xi}{\xi^2} \, \xi\, \frac{d\hat{\sigma}}{\hat{z} d\hat{z} d\Omega} \, d_{\rm mod,red,1} (\xi,M_h) 
+ \text{p.s.} 
+ \order{ \alpha_s} \no
&=z \int_z^1 \frac{d\xi}{\xi^2} \, \xi\, \frac{4 \pi \alpha^2}{3 Q^2}\, \delta (\xi - z) \, z \, d_{\rm mod,red,1} (\xi,M_h) 
+ \text{p.s.} 
+ \order{ \alpha_s} \no
&=\frac{4 \pi \alpha^2}{3 Q^2} \, z\, d_{\rm mod,red,1} (z,M_h) 
+ \text{p.s.} 
+ \order{ \alpha_s} \no
&= z \sigma_0 d_\text{mod,red,1}(z,M_h) 
+ \text{p.s.} 
+ \order{\alpha_s} \, , \label{e.dzcrosssectionndihadron2p}
\end{align}
again with an extra overall factor of $z$ on the right-hand side relative to the normal expression. We could recover the standard expression on the last line by absorbing the $z$ into a redefinition of $d_\text{mod,red,1}(z,M_h)$, but this just converts the definition back into the original one (for $\xi = z$). For a more detailed comparison with standard definition and the definition from Ref.~\cite{Pitonyak:2023gjx}, see again \aref{JAM-Pavia2}.

\section{Comments and discussion}
\label{s.end}

The purpose of this paper is to reaffirm and clarify the validity of prior phenomenological applications of dihadron (or multihadron) fragmentation functions that used a standard operator definition, but extended to include small-mass multihadrons as the observed final state. To this end, we have considered the semi-inclusive production of ``$n$-hadrons'' from $e^+ e^-$ annihilation (SIA) and we have retraced the kinematical steps involved in separating the unfactorized 
cross section into a hard partonic cross section and a fragmentation function. We have verified
that they are unmodified if the observed final state is a low-invariant-mass multihadron. In particular, the prefactor of the resulting fragmentation function definition is the same for all values of $n$. 

An additional goal is to argue that the parton model interpretation of the standard fragmentation function definition also remains applicable even in the case when the produced object is a small-mass multihadron, at least up to the usual caveats and limitations that apply to number density interpretations for all correlation functions. We argue that an alternative definition that was recently introduced~\cite{Pitonyak:2023gjx} is not consistent with factorization derivations unless hard parts and evolution kernels are allowed to acquire nontrivial dependence on external nonperturbative states. For this reason, we advocate for the continued use of the standard definition rather than the newly suggested one. 

To keep the discussion simple, we have focused on the unpolarized multihadron fragmentation functions, but the results apply equally to polarization sensitive quantities like interference fragmentation functions~\cite{Bianconi:1999cd,Bacchetta:2002ux}. 

The topics addressed in this paper mainly concern the formal theoretical foundations of multihadron fragmentation functions in the context of factorization. It remains unclear to us whether and how these results impact the interpretation of existing phenomenological results at a practical level. For example, Refs.~\cite{Cocuzza:2023vqs} and~\cite{Courtoy:2012ry} appear to use identical expressions for the unpolarized factorization formula that relates the SIA cross section to fragmentation functions (compare Eq.~(2) in Ref.~\cite{Cocuzza:2023vqs} and Eq.~(9) in Ref.~\cite{Courtoy:2012ry}), and the same evolution equations are used in both analyses. Therefore, it appears to us on the surface that the steps for extracting the fragmentation functions phenomenologically should be insensitive to the formal operator definitions that are referenced. 
However, page 15 of Ref.~\cite{Cocuzza:2023vqs} asserts that comparisons between the two extractions are obstructed by the differing operator definitions. We leave it for future work to investigate if and to what extent the use of different formal definitions in these phenomenological extractions limits the ability to compare them. 

In future phenomenological applications it will also be important to map out the approximate kinematical regions where factorization with $n$-hadron fragmentation functions can reasonably be expected to remain valid. For any process where a factorization theorem holds, there can be separate relevant ranges of kinematics where different versions of factorization apply, for example in the target versus current regimes of semi-inclusive deep inelastic scattering. 
When the observed final state is an $n$-hadron system, the classification of regions becomes more complicated than in the single hadron case because the invariant mass $M_h$ of the system is no longer fixed and the final state requires a larger set of kinematical variables to fully characterize. In this paper, therefore, we have included discussions about the accuracy of approximations for the case of an observed $n$-hadron final state 
in semi-inclusive $e^+ e^-$ annihilation in somewhat greater detail than is typical. 

While the basic definitions of multihadron fragmentation functions discussed here are unmodified from the single hadron case, further refinements to the treatment of factorization in multihadron cases are needed, and some directions for further work are implied by the discussions provided in this paper. 

As noted, for example, in Refs.~\cite{deFlorian:2003cg,Ceccopieri:2007ip,Zhou:2011ba}, when $\Lambda_{\rm QCD}^2 \ll M_h^2$ the dihadron fragmentation function needs to be refactorized into single hadron fragmentation functions. Moreover, the integrals over $M_h^2$ in the reduced fragmentation functions are ultraviolet divergent, and this leads to extra terms in the collinear evolution equations~\cite{Ceccopieri:2007ip}. Furthermore, when $M_h^2 \approx Q^2$ the factorization formula for the dihadron cross section into dihadron fragmentation functions fails, and one must match the description to a formula where the two hadrons are in entirely separate jets, or where multiple partons fragment.

Indeed, it is generally not straightforward in physical applications to separate fragmentation of a single parton into two hadrons (the mechanism addressed in this paper) from mechanisms where two partons each separately fragment into a hadron. In the intermediate region where $\Lambda_{\rm QCD}^2 \ll M_h^2 \ll Q^2$, it is likely that a resummation of powers of $\ln (M_h^2/Q^2)$ is helpful. The different regions of factorization need to be matched to one another. We intend to discuss this and related aspects of multihadron fragmentation functions further in future work.  

\vskip 0.3in
\acknowledgments
T. Rogers and T. Rainaldi were supported by the U.S. Department of Energy, Office of Science, Office of Nuclear Physics, under Award Number DE-SC0024715. This work
was also supported by the DOE Contract No. DE- AC05-06OR23177, under which 
Jefferson Science Associates, LLC operates Jefferson Lab. 
A. Courtoy was supported  by the UNAM Grants No. DGAPA-PAPIIT IN111222 and IN102225, and CONAHCyT--Ciencia de Frontera 2019 No.~51244 (FORDECYT-PRONACES). We thank J. Osvaldo Gonzalez-Hernandez for useful discussions. 

\appendix 

\section{Momentum components in different variables}
\label{a.momentum}

With variable choices A, the lab frame components are
\begin{align}
p_{h_1}^0 &{}= \frac{z^2 Q^2 (1 + \zeta)^2 + 4(m^2 + \Tscsq{R}{H}) }{4 z (1 + \zeta) Q} \, ,  \\
p_{h_1}^x &{}= \sin(\theta) \cos(\phi) \frac{z^2 Q^2 (1 + \zeta)^2 -4 (m^2 + \Tscsq{R}{H})}{4 z (1 + \zeta) Q }+ \Tsc{R}{H} \cos (\theta ) \cos (\phi ) \cos \left(\phi_{R,H}\right)- \Tsc{R}{H} \sin (\phi ) \sin \left(\phi_{R,H}\right)  \, , \\
p_{h_1}^y &{}= -\sin(\theta) \sin(\phi) \frac{z^2 Q^2 (1 + \zeta)^2 -4 (m^2 + \Tscsq{R}{H})}{4 z (1 + \zeta) Q }- \Tsc{R}{H} \cos (\theta ) \sin (\phi) \cos \left( \phi_{R,H} \right)- \Tsc{R}{H} \cos (\phi) \sin \left(\phi_{R,H}\right) \, , \\
p_{h_1}^z &{}= \cos (\theta )\frac{z^2 Q^2 (1 + \zeta)^2 -4 (m^2 + \Tscsq{R}{H})}{4 z (1 + \zeta) Q }- \Tsc{R}{H} \sin (\theta ) \cos \left(\phi_{R,H}\right) \, , \\ 
p_{h_2}^0 &{}= \frac{z^2 Q^2 (1 - \zeta)^2 + 4(m^2 + \Tscsq{R}{H}) }{4 z (1 - \zeta) Q} \, ,  \\
p_{h_2}^x &{}= \sin(\theta) \cos(\phi) \frac{z^2 Q^2 (1 - \zeta)^2 -4 (m^2 + \Tscsq{R}{H})}{4 z (1 - \zeta) Q }- \Tsc{R}{H} \cos (\theta ) \cos (\phi ) \cos \left(\phi_{R,H}\right)+ \Tsc{R}{H} \sin (\phi ) \sin \left(\phi _R\right)  \, , \\
p_{h_2}^y &{}= -\sin(\theta) \sin(\phi) \frac{z^2 Q^2 (1 - \zeta)^2 -4 (m^2 + \Tscsq{R}{H})}{4 z (1 - \zeta) Q }+ \Tsc{R}{H} \cos (\theta ) \sin (\phi ) \cos \left(\phi_{R,H}\right)+ \Tsc{R}{H} \cos (\phi ) \sin \left(\phi_{R,H}\right) \, , \\
p_{h_2}^z &{}= \cos (\theta )\frac{z^2 Q^2 (1 - \zeta)^2 -4 (m^2 + \Tscsq{R}{H})}{4 z (1 - \zeta) Q }+ \Tsc{R}{H} \sin (\theta ) \cos \left(\phi_{R,H}\right) \,
 . 
\end{align}
With variables choices B, the components of hadron momenta in the laboratory frame 
are
\begin{align}
p_{h_1}^0 &{}= \frac{z_1 Q}{2} \, ,  \\
p_{h_1}^x &{}= \frac{z_1 Q}{2} \left(\sin (\theta ) \cos (\phi ) \sqrt{1 - \frac{\Tscsq{l}{}+4 m^2}{z_1^2 Q^2}}+\frac{\Tsc{l}{}}{z_1 Q} \cos (\theta ) \cos (\phi ) \cos \left(\phi
   _T\right)-\frac{\Tsc{l}{}}{z_1 Q} \sin (\phi ) \sin \left(\phi _T\right)\right)  \, , \\
p_{h_1}^y &{}= \frac{z_1 Q}{2} \left(-\sin (\theta ) \sin (\phi ) \sqrt{1 - \frac{\Tscsq{l}{}+4 m^2}{z_1^2 Q^2}}-\frac{\Tsc{l}{}}{z_1 Q} \cos (\theta ) \sin (\phi ) \cos \left(\phi
   _T\right)-\frac{\Tsc{l}{}}{z_1 Q} \cos (\phi ) \sin \left(\phi _T\right)\right) \, , \\
p_{h_1}^z &{}= \frac{z_1 Q}{2} \left(\cos (\theta ) \sqrt{1 - \frac{\Tscsq{l}{}+4 m^2}{z_1^2 Q^2}}-\frac{\Tsc{l}{}}{z_1 Q} \sin (\theta ) \cos \left(\phi _T\right)\right) \, , \\ 
p_{h_2}^0 &{}= \frac{z_2 Q}{2} \, ,  \\
p_{h_2}^x &{}= \frac{z_2 Q}{2} \left(\sin (\theta ) \cos (\phi ) \sqrt{1 - \frac{\Tscsq{l}{}+4 m^2}{z_2^2 Q^2}}-\frac{\Tsc{l}{}}{z_2 Q} \cos (\theta ) \cos (\phi ) \cos \left(\phi
   _T\right)+\frac{\Tsc{l}{}}{z_2 Q} \sin (\phi ) \sin \left(\phi _T\right)\right) \, ,   \\
p_{h_2}^y &{}= \frac{z_2 Q}{2} \left(-\sin (\theta ) \sin (\phi ) \sqrt{1 - \frac{\Tscsq{l}{}+4 m^2}{z_2^2 Q^2}}+\frac{\Tsc{l}{}}{z_2 Q} \cos (\theta ) \sin (\phi ) \cos \left(\phi
   _T\right)+\frac{\Tsc{l}{}}{z_2 Q} \cos (\phi ) \sin \left(\phi _T\right)\right) \, , \\   
p_{h_2}^z &{}= \frac{z_2 Q}{2} \left(\cos (\theta ) \sqrt{1 - \frac{\Tscsq{l}{}+4 m^2}{z_2^2 Q^2}}+\frac{\Tsc{l}{}}{z_2 Q} \sin (\theta ) \cos \left(\phi _T\right)\right) \, . 
\end{align}
Here, $\T{l}{}$ is the transverse component of $\vect{l}$ in the hadron frame, and $\phi_T$ is its hadron frame azimuthal angle. The three-momenta have magnitudes
\begin{equation}
| \vect{p}_{h_1} | = \frac{z_1 Q}{2} \sqrt{1 - \frac{4 m^2}{z_1^2 Q^2}} \, , \qquad | \vect{p}_{h_2} | = \frac{z_2 Q}{2} \sqrt{1 - \frac{4 m^2}{z_2^2 Q^2}} \, \; .
\end{equation}

\section{Operator definitions of fragmentation functions in Refs.~\cite{Pitonyak:2023gjx,Cocuzza:2023vqs} as compared with 
Ref.~\cite{Bacchetta:2003vn}
}
\label{a.JAM-Pavia2}

To make a more complete comparison between the correlation functions defined in Refs.~\cite{Pitonyak:2023gjx,Cocuzza:2023vqs} and those in 
Ref.~\cite{Bacchetta:2003vn}, 
it is helpful to translate the notation of these earlier papers to that of \srefs{setup}{dihadronexample} of this paper. 
Momentum labels in 
Ref.~\cite{Bacchetta:2003vn}
mostly match those used throughout this paper, with 
$P_h^+ \to p_h^+ \, ,$ 
and other momenta that are capitalized in
Ref.~\cite{Bacchetta:2003vn}
are lower case here. The variables $R$, $\zeta$, and $\Tsc{R}{}$ are defined in the same way as in this paper. In 
Ref.~\cite{Bacchetta:2003vn}, 
the observed hadron moves along the $-z$ axis, so expressions in 
\srefs{setup}{dihadronexample} need to have their plus and minus components reversed in order to be compared. Reference~\cite{Bacchetta:2003vn} 
defines a so-called quark-quark correlator (see 
Eq.(5) of Ref.~\cite{Bacchetta:2003vn}), 
\begin{equation}
\Delta (k, P_h, R) = \sum\!\!\!\!\!\!\!\!\int_{_X} \; \int \frac{\text{d}^4x}{(2\pi)^4}\, e^{i k \cdot x} \, \langle 0 | \psi (x) | P_h, R, X \rangle \langle P_h, R, X | \bar{\psi} (0) | 0 \rangle \; ,
\label{e.DeltaTR}
\end{equation}
which is $1/(2 \pi)^4$ times the momentum space Feynman diagrams of the outgoing quark in \fref{Factorized}. Thus, it is related to $J_1(k,\left\{p_h\right\})$ of \sref{fact} by means of
\begin{equation}
\label{e.Delta2J}
\Delta (k, P_h, R) = \frac{1}{(2\pi)^4} J_1(k,\left\{p_h\right\}) \, .
\end{equation}
By integrating on the light-cone suppressed component $k^+$
and making explicit the dependence from the gauge link operator, the quark-quark correlator becomes (see Eq.~(8) of Ref.~\cite{Bacchetta:2003vn})
\begin{eqnarray}
\Delta (z, \bm{k}_T, R) &= &\int \text{d}k^+ \, \Delta (k, P_h, R) \vert_{k^- = P_h^-/z} \nonumber \\
&= &\sum\!\!\!\!\!\!\!\!\int_{_X} \; \int \frac{\text{d}x^+ \diff[2]{\bm{x}_T}}{(2\pi)^3}\, e^{i k \cdot x} \, \langle 0 | U(\infty, x) \, \psi (x) | P_h, R, X \rangle \langle P_h, R, X | \bar{\psi} (0) \, U(0, x) | 0 \rangle \vert_{x^- = 0} \; . 
\label{e.DeltakTTR}
\end{eqnarray} 
Equation~\eqref{e.DeltakTTR} is identical to the definition in Eqs.(1) and (2) of Ref.~\cite{Pitonyak:2023gjx}, after  expressing it in variables $P_h,R$ rather than $P_1, P_2$ and defining $z=z_1+z_2, \, \zeta = (z_1 - z_2)/z$.

Explicit components of $k$ are in the hadron frame, i.e. $k^+ \to k^+_H$, etc., in the notation of this paper. 
Reference~\cite{Bacchetta:2003vn}
restricts the analysis to the zeroth-order parton model level so that the kinematical $z$ can be identified with the momentum fraction $\xi$. Translating slightly into the language and notation of this paper, the more general way of expressing the operator in \eref{DeltakTTR} would be to translate the equation in the following way: 
\begin{eqnarray}
\Delta (\zetasym, \bm{k}_T, R) &= &\int \text{d}k^+ \, \Delta (k, P_h, R) \vert_{k^- = P_h^-/\zetasym} \nonumber \\
&= &\sum\!\!\!\!\!\!\!\!\int_{_X} \; \int \frac{\text{d}x^+ \diff[2]{\T{x}{}}}{(2\pi)^3}\, e^{i k \cdot x} \, \langle 0 | U(\infty, x) \, \psi (x) | P_h, R, X \rangle \langle P_h, R, X | \bar{\psi} (0) \, U(0, x) | 0 \rangle \vert_{x^- = 0} \; . 
\label{e.DeltakTTR2}
\end{eqnarray} 
Using the parameterization of momenta listed in Eqs.(2,3) of~\cite{Bacchetta:2003vn}, we can rewrite the dependence of the quark-quark correlator as $\Delta (\xi, \zeta, \bm{k}_T, M_h^2, \phi_R)$, where $M_h^2 = P_h^2$ is the dihadron pair invariant mass and $\phi_R$ is the azimuthal angle of $\bm{R}_T$ with respect to the reaction plane. 

The unpolarized fragmentation function $D_1$ can be extracted from the quark-quark correlator through the following projection (see Eq.(16) of Ref.~\cite{Bacchetta:2003vn} or, equivalently, 
Eqs.(11,19) of Ref.~\cite{Bianconi:1999cd}):
\begin{align}
D_1 (\zetasym, \zeta, \bm{k}_T, M_h^2) &= \frac{1}{4 \zetasym} \int \text{d}k^+ \, \mbox{Tr} \left[ \Delta (k, P_h, R) \, \gamma^- \right]\vert_{k^- = P_h^-/\zetasym} \; \no
&= \frac{1}{4 \zetasym} \int \frac{\text{d}k^+}{(2 \pi)^4} \, \mbox{Tr} \left[ J_1(k,\left\{p_h\right\}) \, \gamma^- \right]\vert_{k^- = P_h^-/\zetasym} \, .
\label{e.D1kTTR} 
\end{align}
The factor of $1/(4\zetasym)$ follows from the same factorization-based reasoning as in \sref{fact} and \sref{ffs} of this paper. On the second line, we have applied the notational translation in \eref{Delta2J}, which shows that $D_1 (\zetasym, \zeta, \bm{k}_T, M_h^2)$ matches the definition we gave for $d(\zetasym,-\zetasym \T{k}{H},\left\{ p_h \right\})$ earlier in this paper as it appears in \eref{tmdgraphical}.
By further integrating upon the parton transverse momentum, we get 
\begin{eqnarray}
D_1 (\xi, \zeta, M_h^2) &= &\zetasym^2 \int \diff[2]{\T{k}{}} \, D_1 (\zetasym, \zeta, \bm{k}_T, M_h^2) \nonumber \\
&= &\frac{\zetasym}{4} \int \diff[2]{\T{k}{}} \int \text{d}k^+ \, \mbox{Tr} \left[ \Delta (k, P_h, R) \, \gamma^- \right]\vert_{k^- = P_h^-/\zetasym} \nonumber \\
&= &\frac{\zetasym}{4} \int \frac{\diff{k}^+ \diff[2]{\T{k}{}} }{(2 \pi)^4} \, \mbox{Tr}\, \left[ J_1(k,\left\{p_h\right\}) \, \gamma^- \right] \, . 
\label{e.D1collTR}
\end{eqnarray}
On the last line we have again made the translation in \eref{Delta2J} to emphasize consistency with the definition of $d\parz{\zetasym,\left\{ p_h \right\}}$ in \eref{collinearoperator} of this paper. To summarize the above, the dihadron fragmentation functions of Ref.~\cite{Bacchetta:2003vn} 
and related articles are defined in the same way as in \sref{ffs} of this paper. 

To change variables and compare these expressions with those of 
Ref.~\cite{Pitonyak:2023gjx}, it is necessary to consider the integration over some regions of phase space $V$ where factorization is reasonable. In preparation for this, note from \eref{diffzomega1} and \eref{varchange1} the following variable changes in integrals over $P_1$ and $P_2$, 
\begin{align}
& \int \frac{\diff[3]{\vect{P}_1} \diff[3]{\vect{P}_2}}{2 E_1 (2 \pi)^3 2 E_2 (2 \pi)^3} \parz{1 + \cos^2 \theta} \ldots \no
& = \frac{Q^2}{32 (2\pi)^6} \int \diff{\Omega} z \diff{z} \diff{M_h} \diff{\zeta} \diff{\phi_{R}} M_h \parz{1 + \cos^2 \theta} \ldots = \frac{1}{3} \frac{16 \pi Q^2}{32 (2\pi)^6} \int  z \diff{z} \diff{M_h} \diff{\zeta} \diff{\phi_{R}} M_h \ldots
\label{e.inttrans1}
\end{align}
where the ``$\ldots$'' only has angle-independent factors. 
Another useful variable transformation is 
\begin{align}
& \int \frac{\diff[3]{\vect{P}_{p,1}} \diff[3]{\vect{P}_{p,2}}}{2 E_{p,1} (2 \pi)^3 2 E_{p,2} (2 \pi)^3} 4 z_1 z_2 (2 \pi)^6 \ldots = \int \diff{z_1} \diff[2]{\T{P}{1,p}} \diff{z_2} \diff[2]{\T{P}{2,p}} \ldots \no
& =  \int \diff{z} \diff{M_h} \diff{\zeta} \diff{\phi_{R}} \diff[2]{\T{k}{}} (1 - \zeta^2)\frac{z^3}{8} M_h  \ldots \, .
\label{e.inttrans2}
\end{align}
Integrals of this latter type should be interpreted with care since the momentum components of the hadrons in the parton frame do not directly correspond to a physically measurable momentum. Rather, the parton frame moves as the internal integrals over partonic $\T{k}{}$ are carried out. Parton frame variables are ill-defined for describing the differential cross section because parton momenta are internal integration variables. 

Because $D_1 (\zetasym, \zeta, \bm{k}_T, M_h^2)$ matches the definition from \sref{ffs}, the factorization derivation relates \eref{D1collTR} to a zeroth-order parton model cross section via \eref{fact_formula_zeroth}:
\begin{align}
\frac{1}{\sigma_0} & \int_V \frac{\diff[3]{\vect{P}_1} \diff[3]{\vect{P}_2}}{2 E_1 (2 \pi)^3 2 E_2 (2 \pi)^3} \parz{1 + \cos^2 \theta} \parz{\frac{2 \alpha^2 (2 \pi)^3}{z Q^4}} D_1 (z, \zeta, M_h^2) = \langle N_2 \rangle_V + \text{p.s.} + \order{\alpha_s} \no
&\to \frac{Q^4}{8 (2\pi)^6} \int_V  M_h \diff{z} \diff{M_h} \diff{\zeta} \diff{\phi_{R}} \parz{\frac{2 (2 \pi)^3}{Q^4}} D_1 (z, \zeta, M_h^2) = \langle N_2 \rangle_V + \text{p.s.} + \order{\alpha_s} \no
&\to \frac{1}{16 \pi^2} \int_V  M_h \diff{z} \diff{M_h} \diff{\zeta} D_1 (z, \zeta, M_h^2) = \langle N_2 \rangle_V + \text{p.s.} + \order{\alpha_s} \, .
\label{e.xsectionothernot}
\end{align}
We have divided the semi-inclusive cross section on the left-hand side by the total cross section $\sigma_0$ to convert it into a dihadron multiplicity $\langle N_2 \rangle_V$. The $V$ subscripts are to emphasize the restriction on the integrals to the factorization region. Just as in \eref{extended}, we may next define
\begin{equation}
D_{1,\text{red}} (z,M_h) \equiv \int_{-\zeta_\text{max}}^{\zeta_\text{max}} \diff{\zeta}{} \int_0^{2 \pi} \diff{\phi_{R}}{}  \frac{M_h}{32 \pi^3} D_1 (z, \zeta, M_h^2) = 
\int_{-\zeta_\text{max}}^{\zeta_\text{max}} \diff{\zeta}{}   \frac{M_h}{16 \pi^2} D_1 (z, \zeta, M_h^2) .
\label{e.reducedapp}
\end{equation}
So, 
\begin{equation}
\int_V \diff{z} \diff{M_h} D_{1,\text{red}} (z,M_h) = \langle N_2 \rangle_V + \text{p.s.} + \order{\alpha_s} \, .
\end{equation}
If we use $\langle N_2 \rangle_V = \sigma_V^{\text{inclusive}}/\sigma_0$ this is just the semi-inclusive differential cross section,  
\begin{equation}
\frac{\diff{\sigma}}{\diff{z} \diff{M_h}} = \sigma_0 D_{1,\text{red}} (z,M_h) + \text{p.s.} + \order{\alpha_s} = \frac{4 \pi \alpha^2}{3 Q^2} D_{1,\text{red}} (z,M_h) + \text{p.s.} + \order{\alpha_s} \, , \label{e.crosssectionapp}
\end{equation}
thus reproducing \eref{dzcrosssectionndihadron} of this paper but now in the notation of Ref.~\cite{Bacchetta:2003vn}
and related papers.

An alternative way that one might attempt to obtain a dihadron multiplicity is to instead try to enforce the number sum rule of Ref.~\cite{Pitonyak:2023gjx} on the basic definition in \eref{D1kTTR}: 
\begin{equation}
\int \diff{z_1} \diff[2]{\T{P}{1,p}} \diff{z_2} \diff[2]{\T{P}{2,p}} D_1 (z, \zeta, \bm{k}_T, M_h^2) = \langle N_2 \rangle_V \, , \label{e.wrongmult}
\end{equation}
where, in the spirit of remaining in a zeroth-order parton model approximation, we have fixed $\zetasym = z$. 
Using \eref{inttrans2}, this would give 
\begin{align}
&\int \diff{z} \diff{M_h} \diff{\zeta} \diff{\phi_{R}} (1 - \zeta^2)\frac{z^3}{8} M_h \diff[2]{\T{k}{}} D_1 (z, \zeta, \bm{k}_T, M_h^2) \, \no
&= \pi \int \diff{z} \diff{M_h} \diff{\zeta}  (1 - \zeta^2)\frac{z}{4} M_h  D_1 (z, \zeta, M_h^2) = \langle N_2 \rangle_V \, ,
\label{e.sumrulepcmps}
\end{align}
but this clearly contradicts the last line of  \eref{xsectionothernot}, which was obtained by working directly with the cross section, especially if we wish to interpret $z$ as a momentum fraction and not a process dependent external kinematical variable. Hence, there is the appearance of a paradox.\footnote{The presence of a contradiction relies on an assumption that the physical multiplicity $\langle N_2 \rangle_V$ in \eref{xsectionothernot} is to be identified approximately with the ``sum rule'' quantity on the right-hand side of \eref{wrongmult}, labeled by the same symbol.}

In order to force the last line of \eref{sumrulepcmps} 
to match the last line of \eref{xsectionothernot}, 
one would need to redefine the  
fragmentation function used in \eref{sumrulepcmps} and replace it with a new fragmentation function $D_1' (z, \zeta, M_h^2)$ by setting the integrands of \eref{sumrulepcmps} and \eref{xsectionothernot} equal: 
\begin{align}
\frac{\pi z}{4} (1 - \zeta^2) D_1' (z, \zeta, M_h^2) &= \frac{1}{16 \pi^2} D_1 (z, \zeta, M_h^2) \no
\to \quad D_1' (z, \zeta, M_h^2) &= \frac{z}{16 \pi^3 z_1 z_2} D_1 (z, \zeta, M_h^2)
\, . \label{e.pcmpsfinal}
\end{align}
Using \eref{D1collTR}, we then get
\begin{align}
D_1' (z, \zeta, M_h^2) &= z^2 \int \diff[2]{\T{k}{}} \left[ \frac{1}{64 \pi^3 z_1 z_2} \int \text{d}k^+ \, \mbox{Tr} \left[ \Delta (k, P_h, R) \, \gamma^- \right]\vert_{k^- = P_h^-/z} \right] \no
&= \int \diff[2]{\T{k}{}} \frac{1}{16 \pi^3 (1 - \zeta^2)} \int \text{d}k^+ \, \mbox{Tr} \left[ \Delta (k, P_h, R) \, \gamma^- \right]\vert_{k^- = P_h^-/z} \, \no
&\equiv  \int \diff[2]{\T{k}{}} D'_1 (z, \zeta,\T{k}{}, M_h^2) \, , 
\label{e.pcmpsfinal2}
\end{align} 
where the quantity in braces on the first line is \eref{ffdefmod}, i.e. the definition from Ref.~\cite{Pitonyak:2023gjx}, with $\xi$ fixed to $z$, and in the last line we have adopted the same convention of Ref.~\cite{Pitonyak:2023gjx} to relate collinear and transverse-momentum dependent fragmentation functions via an integral over $\T{k}{}$ rather than over $z^2 \T{k}{}$ and $\phi_R$. But $D_1 (z, \zeta, M_h^2)$ and $D'_1 (z, \zeta, M_h^2)$ cannot both be the basic fundamental definition of the fragmentation function once $\zetasym \neq z$.

Despite the contradiction between \eref{xsectionothernot} and \eref{sumrulepcmps}, let us continue to study the properties of the new primed fragmentation functions under various variable transformations and notation changes. It may be desired to always define fragmentation functions in the form of densities in the exact arguments used to parametrize them, and this is readily done by grouping Jacobian factors together each time variables are changed and then defining new functions. We can do this, for example, by switching to integrals over  $R_T^2$ rather than $M_h^2$.  Of course, all the expressions above can be written in terms of integrals over $R_T^2$ rather than $M_h^2$ if, using \eref{RT2Mh}, we simply make the replacement 
\begin{align}
&{} \int M_h \diff{M_h}{} \ldots = \int \frac{1}{2} \diff{M_h^2}{} \ldots =  
\int \frac{2}{1 - \zeta^2} \diff{\Tscsq{R}{}} \ldots 
\end{align}
wherever an integral over $M_h$ appears. All fragmentation functions are made to be explicit functions of $R_T^2$
if, again using \eref{RT2Mh}, we write 
\begin{align}
 D_1 (z, \zeta, R_T^2) &\equiv D_1 \parz{z, \zeta, M_h^2(R_T^2,\zeta)} \label{e.RT2Mh2} \\
 D_1 (z, \zeta, \bm{k}_T, R_T^2) &\equiv D_1 \parz{z, \zeta, \bm{k}_T, M_h^2( R_T^2,\zeta)} \, .
\end{align}
Then, \eref{sumrulepcmps} becomes
\begin{align}
&\int \diff{z} 
\diff{\zeta} \diff{\phi_{R}} \diff{R_T^2} \frac{z^3}{4} \diff[2]{\T{k}{}} D_1 (z, \zeta, \bm{k}_T, R_T^2) \, \no
&= \int \diff{z} \diff{\zeta} 
\diff[2]{\T{R}{}} \frac{z}{2}  D_1 (z, \zeta, R_T^2) = \langle N_2 \rangle_V \, .
\label{e.sumrulepcmpsRT}
\end{align}
The transition to primed functions is exactly the same for functions of $R_T^2$
as for the functions of $M_h^2$. So, rewriting \eref{sumrulepcmpsRT} for $D_1'(z,\zeta, R_T^2)$ and applying a definition analogous to the last line of \eref{pcmpsfinal2}, we get
\begin{align}
\int \diff{z} \diff{\zeta} \diff[2]{\T{R}{}} \parz{\diff[2]{\T{k}{}} \frac{z}{2}   D'_1 (z, \zeta,\T{k}{},  R_T^2)} = \langle N_2 \rangle_V \, .
\label{e.sumrulepcmps2}
\end{align}
\vspace{0.1cm}
Noting the integrand, we can define
\begin{align}
 D_1^{\text{JAM}} (z, \zeta,R_T^2) \equiv \int  \diff[2]{\T{k}{}} \frac{z}{2}   D'_1 (z, \zeta, \T{k}{}, R_T^2) = \frac{z}{32 \pi^3 (1 - \zeta^2)} \int \diff[2]{\T{k}{}} \int \text{d}k^+ \, \mbox{Tr} \left[ \Delta (k, P_h, R) \, \gamma^- \right]\vert_{k^- = P_h^-/z} \, ,
\label{e.sumrulepcmpsprimenew} 
\end{align}
 which matches Eq.~(15) of Ref.~\cite{Pitonyak:2023gjx}. If we swap the fragmentation function for the primed fragmentation function and use \eref{RT2Mh2} then the second line of \eref{sumrulepcmps} becomes 
\begin{align}
&\pi \int \diff{z} \diff{M_h} \diff{\zeta}  (1 - \zeta^2)\frac{z}{4} M_h  D_1' (z, \zeta, M_h^2) 
= \pi \int \diff{z} \diff{M_h} \diff{\zeta}  (1 - \zeta^2)\frac{z}{4} M_h  D_1' (z, \zeta, R_T^2) \no
&= \pi \int \diff{z} \diff{M_h} \diff{\zeta}  (1 - \zeta^2)\frac{z}{4} M_h 
\int \diff[2]{\T{k}{}} 
D'_1 (z, \zeta,\T{k}{}, R_T^2) 
= \int \diff{z} \diff{M_h} \left( \frac{\pi}{2} \int \diff{\zeta}{} M_h (1-\zeta^2) D_1^{\text{JAM}} (z, \zeta,R_T^2) \right) \no
&\equiv \int \diff{z} \diff{M_h} D_1^{\text{JAM}} (z,M_h) =  \langle N_2 \rangle_V \, ,
\label{e.sumrulepcmpsprime}
\end{align}
and the $D_1^{\text{JAM}} (z,M_h)$ in the last line then matches Eq.(17) of Ref.~\cite{Pitonyak:2023gjx}. Note that if we work strictly with the $\xi = z$ assumption, then the $(1 - \zeta^2)$ that was absorbed into the definition of $D_1^{\text{JAM}} (z,M_h)$ simply converts it back to the original definition in \eref{D1collTR} and the remaining factors make it a $\xi = z$ version of the reduced fragmentation function in \eref{extended} (or \eref{reducedapp} above).

To summarize, we have shown how, starting from the expressions and notation in Ref.~\cite{Bacchetta:2003vn}, 
one may modify the operator definitions of the fragmentation functions that do follow from factorization into ``primed definitions" that instead preserve \eref{wrongmult} via \eref{pcmpsfinal} and reproduce the newly suggested operator definition in Eq.(4) of Ref.~\cite{Pitonyak:2023gjx}. We then connected these modified operator definitions to the $D_1^{\text{JAM}} (z, \zeta,R_T^2)$ and $D_1^{\text{JAM}} (z,M_h)$ functions in \erefs{sumrulepcmpsprimenew}{sumrulepcmpsprime} above by reabsorbing various factors to recover the other expressions from Ref.~\cite{Pitonyak:2023gjx}. 
In retracing the steps above, we are able to see that the fundamental definitions obtained from factorizing the cross section differ in a concrete way from the modified definitions that are obtained by imposing a sum rule like \eref{wrongmult}. 
 
The resolution to the paradox is that \eref{wrongmult} is not a valid relation for the basic definition, as explained in the main text and in Refs.~\cite{Rogers:2024zvi,Collins:2023cuo}. \\

\section{Effect of extra prefactors on evolution}
\label{a.yukawa}

First, consider the definition with the standard prefactor for both the single and dihadron cases and use the Yukawa theory toy model discussed in \sref{alternative}. 
For the zeroth order bare scalar-in-scalar fragmentation function, 
\begin{equation}
d^{[\order{1}]}_{0,s/s}(\zetasym) = k^+ \zetasym^3 \int \frac{\diff{k^-} \diff[2]{\T{k}{}}}{(2 \pi)^4} (2 \pi)^4 \delta^{(4)}(k - \hv{}) = \delta(1-\zetasym) \, .
\end{equation}
For the zeroth order bare dihadron-in-scalar fragmentation function, 
\begin{equation}
d_{0,ss/s}^{[\order{1}]}(\zetasym) = \frac{\lambda^2}{M_h^4} k^+ \zetasym^3 \int \frac{\diff{k^-} \diff[2]{\T{k}{}}}{(2 \pi)^4} (2 \pi)^4 \delta^{(4)}(k - \hv{}) = \frac{\lambda^2}{M_h^4} \delta(1-\zetasym) \, .
\end{equation}
At $\order{g^2}$ in the Yukawa theory, the single hadron and dihadron fragmentation functions that follow from their operator definitions also satisfy
\begin{equation}
d_{0,ss/q}^{[\order{g^2}]}(\zetasym) = \frac{\lambda^2}{M_h^4} d_{0,s/q}^{[\order{g^2}]}(\zetasym) \, .
\end{equation}
The $\msbar$ quark-in-scalar renormalization factor expanded in perturbation theory is
\begin{equation}
Z_{qj'}(\zetasym) = \delta_{qj'} \delta(1- \zetasym) + g^2 \frac{S_\epsilon}{\epsilon} \delta_{qj'} Z^{[\order{g^2}]}_{j'/s}(\zetasym) + \order{g^4} \, ,
\end{equation}
where the dimension of spacetime is $4 - 2\epsilon$ and $S_\epsilon = (4 \pi)^\epsilon/\Gamma(1-\epsilon)$.
Expanded in perturbation theory, \eref{renorm} is 
\begin{equation}
\lim_{\epsilon \to 0} \sum_{j'} \int_z^1 \frac{\diff{\zetasym'}}{\zetasym'} \left[ d_{0,s/q}^{[\order{g^2}]} (\zetasym/\zetasym') \delta(1- \zetasym') + g^2 \frac{S_\epsilon}{\epsilon} Z^{[\order{g^2}]}_{q/s}(\zetasym) d_{0,s/s}^{[\order{1}]}(\zetasym/\zetasym') \right] + \order{g^4}
\end{equation}
and in the dihadron case, 
\begin{equation}
\lim_{\epsilon \to 0} \sum_{j'} \int_z^1 \frac{\diff{\zetasym'}}{\zetasym'} \frac{\lambda^2}{M_h^4} \left[ d_{0,s/q}^{[\order{g^2}]} (\zetasym/\zetasym') \delta(1- \zetasym') + g^2 \frac{S_\epsilon}{\epsilon} Z^{[\order{g^2}]}_{q/s}(\zetasym) d_{0,s/s}^{[\order{1}]}(\zetasym/\zetasym') \right] + \order{g^4} \, . \label{e.direnorm}
\end{equation}
The $Z^{[\order{g^2}]}_{q/s}(\zetasym)$ in the second term needs to ensure the cancellation of the $S_\epsilon/\epsilon$ pole in the first term.
Clearly, then, the same $Z^{[\order{g^2}]}_{q/s}(\zetasym)$ applies to both the single and dihadron cases. The only difference between the two cases is the overall $\frac{\lambda^2}{M_h^4}$ in the dihadron case that arises from the dihadron/scalar vertex.

Now consider the consequence of modifying the prefactor of the dihadron fragmentation function definition by a factor $\mathcal{N}(\zetasym) = \zetasym^\nu$ (with $\nu \neq 0$) relative to the single hadron fragmentation function. Then in the above we have the replacements, 
\begin{align}
   d_{0,ss/s}^{[\order{1}]}(\zetasym) \to \; &{} \mathcal{N}(\zetasym) \frac{\lambda^2}{M_h^4} d_{0,s/s}^{[\order{1}]}(\zetasym) = \mathcal{N}(1) \frac{\lambda^2}{M_h^4} d_{0,s/s}^{[\order{1}]}(\zetasym) \qquad \text{and} \no
d_{0,ss/q}^{[\order{g^2}]}(\zetasym/\zetasym') \to \; &{} \mathcal{N}(\zetasym/\zetasym') \frac{\lambda^2}{M_h^4} d_{0,s/q}^{[\order{g^2}]}(\zetasym/\zetasym') \, .
\end{align}
Now \eref{direnorm} becomes
\begin{equation}
\lim_{\epsilon \to 0} \sum_{j'} \int_z^1 \frac{\diff{\zetasym'}}{\zetasym'} \frac{\lambda^2}{M_h^4} \mathcal{N}(1) \left[ \frac{\mathcal{N}(\zetasym)}{\mathcal{N}(1)} d_{0,s/q}^{[\order{g^2}]} (\zetasym/\zetasym') \delta(1- \zetasym') + g^2 \frac{S_\epsilon}{\epsilon} Z^{[\order{g^2}]}_{q/s}(\zetasym) d_{0,s/s}^{[\order{1}]}(\zetasym/\zetasym') \right] + \order{g^4} \, . \label{e.direnorm2}
\end{equation}
Here we have used $\mathcal{N}(\zetasym/\zetasym') = \mathcal{N}(\zetasym)$ because of the $\delta(1 - \zetasym')$. Comparing with \eref{direnorm} shows that the order-$g^2$ correction to the $\msbar$ renormalization factor is modified from the result in the standard definition by a factor of $\mathcal{N}(\zetasym)/\mathcal{N}(1)$ for the dihadron case,
\begin{equation}
\left( Z^{[\order{g^2}]}_{q/s}(\zetasym) \right)_{\text{dihadron}} = \frac{\mathcal{N}(\zetasym)}{\mathcal{N}(1)} \left( Z^{[\order{g^2}]}_{q/s}(\zetasym) \right)_{\text{single hadron}}
\end{equation}
The splitting functions $P_{jj'}$ are obtained by inverting the 
equation for the running of the renormalization factor,
\begin{equation}
\frac{\diff{}{}}{\diff{\ln \mu^2}{}} Z_{jk}(\zetasym) = \sum_{j'} \int \frac{\diff{\zetasym'}{}}{\zetasym'} P_{jj'}(\zetasym') Z_{j'k}(\zetasym/\zetasym') \, .
\end{equation}
Therefore, modifying the prefactor of the operator definition of the multihadron fragmentation function by a nontrivial $\zetasym$-dependent factor necessarily leads to a modification of the evolution kernel $P_{jj'}$ for the multihadron case relative to the single hadron case. 

Note that for treating evolution it is the momentum fraction $\zetasym$ that is relevant and not external kinematical factors of $z$, etc., that might arise from the Jacobians when changing the variables used to parametrize a cross section for a specific process.

The evolution is unaffected if $\mathcal{N}(\zetasym)$ consists only of constant factors like $\pi$ since these cancel from both sides.

\bibliography{bibliography}

\end{document}